\documentclass{article}
\usepackage[english]{babel}
\usepackage{cite}
\usepackage{bbold}

\usepackage[letterpaper,top=2cm,bottom=2cm,left=3cm,right=3cm,marginparwidth=1.75cm]{geometry}

\usepackage{amsmath}
\usepackage{graphicx}
\usepackage[colorlinks=true, allcolors=blue]{hyperref}

\title{Quantum corrections to the Weyl quantization of the classical time of arrival}
\author{Dean Alvin L. Pablico$^1$, and Eric A. Galapon$^2$}

\begin{document}
\maketitle

\begin{abstract}
A time of arrival (TOA) operator that is conjugate with the system Hamiltonian was constructed by Galapon without canonical quantization in [J. Math. Phys. \textbf{45},  3180 (2004)]. The constructed operator was expressed as an infinite series but only the leading term was investigated which was shown to be equal to the Weyl-quantized TOA-operator for entire analytic potentials. In this paper, we give a full account of the said TOA-operator by explicitly solving all the terms in the expansion. We interpret the terms beyond the leading term as the quantum corrections to the Weyl quantization of the classical arrival time. These quantum corrections are expressed as some integrals of the interaction potential and their properties are investigated in detail. In particular, the quantum corrections always vanish for linear systems but nonvanishing for nonlinear systems. Finally, we consider the case of an anharmonic oscillator potential as an example. 
\end{abstract}

\section{Introduction}

Quantum mechanics is one of the most successful quantitative theory ever formulated in describing the physical properties of nature. Its foundations have been tested rigorously and its predictions have been experimentally verified to a high degree of precision and accuracy. However, quantum mechanics is still considered incomplete despite being considered as the triumph of the twentieth-century science. One of the reasons is the lack of formal treatment of time quantities as quantum dynamical observables, a problem known in the literature as the quantum time problem. While there is a rich literature devoted to this problem, a consensus on its resolution has yet to be reached (see, for instance, Refs. \cite{Aharonov1961,Pollak1984,Giannitrapani1997,Peres1980,Hilgevoord2002,Olkhovsky2007,Olkhovsky2008,Muga2009,Bauer2017,Leon2017,Maccone2020,Jurman2021,Allcock1969a,Allcock1969b,Allcock1969,Grot1996,Aharonov1998,Leavens1998,Delgado1998,Baute2000,Wang2007,Muga2000,J.Leon20000,Galapon2001,Galapon2004,Galapon2018,Galapon2009a,Galapon2006,Muga2008,Galapon2002,Halliwell2015,Pollak2017,Sombillo2018,Galapon2004a,Anastopoulos2006,Sombillo2016,Das2021,Galapon2002a,Galapon2005,Galapon2005a,Galapon2008,Caballar2009,Caballar2010,Villanueva2010,Flores2019} and all references therein). 

One facet of the quantum time problem is the quantum time of arrival problem which requires finding the quantum time of arrival distribution of an elementary particle at a given point in the configuration space \cite{Allcock1969a,Allcock1969b,Allcock1969,Grot1996,Aharonov1998,Leavens1998,Delgado1998,Wang2007,Muga2000,J.Leon20000,Galapon2001,Galapon2004,Galapon2018,Galapon2009a,Galapon2006,Baute2000,Muga2008,Galapon2002,Halliwell2015,Pollak2017,Sombillo2018,Galapon2004a,Anastopoulos2006,Sombillo2016,Das2021,Galapon2002a,Galapon2005,Galapon2005a,Galapon2008,Caballar2009,Caballar2010,Villanueva2010,Flores2019}. In classical mechanics, the time of arrival problem is more of a textbook problem. For a structureless particle of mass $\mu$ initially located at a point $(q,p)$ at $t=0$ in the phase space, the corresponding time of arrival $t=T_x(q,p)$ at some point $q(t=T_x)=x$ in the configuration space is given by
\begin{equation}\label{classical}
T_x(q,p)=-\mathrm{sgn}(p)\sqrt{\frac{\mu}{2}}\int_{x}^{q} \, \frac{dq'}{\sqrt{H(q,p)-V(q')}},
\end{equation}
where $V(q)$ is the interaction potential and $H(q,p)$ is the corresponding Hamitonian. Equation (\ref{classical}) is simply derived by inverting the classical equations of motion of the particle. As a classical observable, $T_x(q,p)$ is always finite and real valued in all classically accessible regions in the phase space. It may also have multiple values indicating possible multiple arrivals at the arrival point. Most importantly, it satisfies the classical conjugacy requirement with the Hamiltonian, i.e., it satisfies the canonical Poisson bracket relation 
$\{T_x(q,p),H(q,p)\}=-1$.

The quantum time of arrival problem entails obtaining the quantum image of the classical time of arrival. Following the standard theoretical framework of quantum mechanics, the problem implies finding the appropriate TOA-operator in the underlying Hilbert space of the physical system. The physical contents of the constructed operator can then be extracted by investigating its eigenfunctions, eigenvalues, expectation value and the corresponding probability distribution for arrival time measurements. However, the construction of TOA-operators has been initially ignored due to Pauli who argued that no self-adjoint time operators canonically conjugate to semibounded Hamiltonians exist \cite{Pauli1926}. Pauli's ``theorem" fundamentally forbids the use of the operator formalism in resolving quantum time of arrival problems, strongly suggesting that time quantities should remain as parameters. Nevertheless, it has already been established by one of us that Pauli's no-go theorem does not hold in single Hilbert spaces and thus there is no a priori reason to dismiss the existence of self-adjoint time operators canonically conjugate to a semibounded Hamiltonian \cite{Galapon2002,Galapon2006atime}. In addition, there is a growing consensus that time observables are actually positive operator valued measures (POVMs) in which they can be formally represented by maximally symmetric but not necessarily self-adjoint operators \cite{Giannitrapani1997,Muga2008}. Hence, the construction of time of arrival operators remains a valid and meaningful method in solving quantum time of arrival problems. 

One possible way of constructing time of arrival operators is the canonical quantization of the classical time of arrival  \cite{Aharonov1961,Wang2007,Muga2000,J.Leon20000,Galapon2001,Galapon2004,Galapon2018}. In Ref. \cite{Galapon2004}, quantization was done on the local time of arrival (LTOA) which is the expansion of the classical time of arrival $T_x(q,p)$ about the free particle arrival time at the arrival point $x$. For arrivals at the origin ($x=0$), the LTOA is given by
\begin{equation}\label{ltoa}
\tau(q,p)=-\sum_{k=0}^{\infty}(-1)^k \frac{(2k-1)!!}{k!} \frac{\mu^{k+1}}{p^{2k+1}}\int_{0}^{q}dq' \, \left(V(q)-V(q')\right)^k.
\end{equation}
Using the rigged Hilbert space formulation (RHS) of quantum mechanics, the Weyl quantization of the local time of arrival in coordinate representation is given by
\begin{equation}\label{toadef}
(\hat{\mathrm{T}}\varphi)(q)=\int_{-\infty}^{\infty} dq' \,\langle q|\hat{T}|q'\rangle \varphi(q').
\end{equation}
We formally refer to the above integral operator as the Weyl-quantized TOA-operator. Its kernel $\langle q|\hat{T}|q'\rangle$ assumes the form
\begin{equation}\label{timekernel}
\langle q|\hat{T}|q'\rangle=\frac{\mu}{i \hbar}\, \mathrm{sgn}(q-q')\,T(q,q'),
\end{equation}
where $T(q,q')=T_W(q,q')$ is referred to as the time kernel factor and is given by
\begin{equation}\label{tweyl}
T_W(q,q')=\frac{1}{4}\int_{0}^{q+q'} ds \, {}_0F_{1}\left(;1;\left(\frac{\mu}{2\hbar^2}\right)(q-q')^2\left[V\left(\frac{q+q'}{2}\right)-V\left(\frac{s}{2}\right)\right]\right).
\end{equation}
The factor ${}_0F_{1}(;1;z)$ is a specific hypergeometric function. The Weyl-quantized TOA-operator is considered a legitimate TOA-operator since its eigenfunctions exhibit unitary arrival at the intended arrival point at a time equal to its eigenvalue. This provides a direct connection between the collapse of the particle's wavefunction and its appearance at the arrival point \cite{Galapon2006,Sombillo2016}. It has been applied to specific quantum time of arrival problems including the explicit calculations of the time-of-flight of free neutrons \cite{Galapon2009}, the quantum tunneling time in a single rectangular barrier \cite{Galapon2012}, and the traversal time across a potential well \cite{Pablico2020}; the investigation of the generalized Hartman effect in a multiple barrier system \cite{Sombillo2014}, and the interpretation of the time-energy uncertainty relation \cite{Sombillo2018}.

However, the Weyl-quantized TOA-operator is not the only possible TOA-operator that can represent a TOA observable. The non-commutativity of the position and momentum operators leads to ordering ambiguities. As a consequence, a single classical time of arrival quantity may have an infinite number of quantizations. In Ref. \cite{Galapon2018}, two additional TOA-operators were introduced using Born-Jordan and symmetric quantizations of the classical time of arrival. They also share the same integral form as the Weyl-quantized TOA-operator (\ref{toadef}) but with different kernel factors $T(q,q')$. The Weyl, Born-Jordan, and symmetric quantized TOA-operators lead to the same operator for the free-particle case which satisfies the conjugacy requirement with the free-particle Hamiltonian. On the other hand, the three quantized TOA-operators differ for the non-interacting case and only the Weyl-quantized TOA-operator satisfies the conjugacy requirement but only for linear systems. For nonlinear systems, quantization of the classical time of arrival fails to provide TOA-operators that satisfy the required commutator values. These findings illustrate the well-known presence of obstruction to quantization in important spaces such as the Euclidean space \cite{Gotay1999,Groenewold1946}. This raises a fundamental question as to whether or not canonical quantization correctly and consistently describes experimental measurements of the arrival time of a quantum particle. And even if it is, another problem arises on how to choose which quantization rule represents a specific time of arrival observable.

Aside from the problem of obstruction to quantization, there exists a circularity in quantization due to the fact that quantization presupposes the axioms of classical mechanics to derive quantum observables, and then one invokes the correspondence principle to rederive the classical observables. This, in itself, is problematic because quantum mechanics should be internally coherent and autonomous from classical mechanics and that classical mechanics should just remain as the contraction of quantum mechanics  \cite{Galapon2001}. This issue has always been ignored because canonical quantization works in many of its practical applications. But the presence of obstructions to quantization and the existence of quantum effects with no classical counterparts, such as quantum tunneling, motivate us to look for other method that is independent on the canonical quantization of classical time observables.

To address these issues on quantization, supraquantization was introduced in Ref. \cite{Galapon2004} as a method of constructing time of arrival operators without quantization. The constructed operator is referred to as the supraquantized TOA-operator and is constructed from purely quantum mechanical considerations. In this formalism, the classical time of arrival is just the contraction of the supraquantized TOA-operator in some limit of vanishing $\hbar$ and not the starting point of calculations, in contrast to quantization. It bypasses the issue of obstruction to quantization since the constructed operator is always conjugate with the system Hamiltonian.  

The supraquantized TOA-operator $\hat{\mathrm{T}}$ and its kernel $\langle q|\hat{T}|q'\rangle$ also share the same general form as the quantized ones (\ref{toadef}-\ref{timekernel}). The difference, however, lies in the form of the kernel factor $T(q,q')$. In supraquantization, the kernel factor $T(q,q')$ is determined as the solution of a second order partial differential equation, called the time kernel equation (TKE), with specific boundary conditions rather than as the quantization of the local time of arrival. The time kernel equation is a direct consequence of the conjugacy of the supraquantized TOA-operator with the system hamiltonian $\hat{\mathrm{H}}$ while the boundary conditions to be imposed ensure that the operator has the correct classical limit \cite{Galapon2004}. Supraquantization then translates a specific quantum time of arrival problem into solving the corresponding time kernel equation with the specified boundary conditions. 

An important feature of the time kernel equation is that its solution $T(q,q')$ admits the following expansion for entire analytic potentials 
\begin{equation}\label{kernelexpand}
T(q,q')=T_0(q,q')+T_1(q,q')+T_2(q,q')+...
\end{equation}
so that the kernel $\langle q|\hat{T}|q'\rangle$ of the supraquantized TOA-operator appears to be
\begin{equation}\label{expandkernel}
\langle q|\hat{T}|q'\rangle=\langle q|\hat{T}_0|q'\rangle+\langle q|\hat{T}_1|q'\rangle+\langle q|\hat{T}_2|q'\rangle+...=\left(\frac{\mu}{i\hbar}\right)\mathrm{sgn}(q-q')\sum_{n=0}^{\infty}T_n(q,q').
\end{equation}
Accordingly, the supraquantized TOA-operator itself formally appears as the expansion
\begin{equation}\label{supraexpand}
\hat{\mathrm{T}}=\hat{\mathrm{T}}_0+\hat{\mathrm{T}}_1+\hat{\mathrm{T}}_2+...
\end{equation}
The significance of the above form is unraveled by taking the Weyl-Wigner transform of the kernel $\left\langle q\left|T\right|q'\right\rangle$ of the supraquantized TOA-operator which is given by
\begin{equation}
\begin{split}
\mathcal{T}_\hbar(q,p)&=\int_{-\infty}^{\infty}d\nu \,\left\langle q+\frac{\nu}{2}\left|\hat{T}\right|q-\frac{\nu}{2}\right\rangle \mathrm{e}^{-i \nu p/\hbar}\\
&=\sum_{n=0}^{\infty}\frac{\mu}{i\hbar}\int_{-\infty}^{\infty}d\nu\,\, T_n\left(q+\frac{\nu}{2},q-\frac{\nu}{2}\right)\,\mathrm{sgn}(\nu)\,\mathrm{e}^{-ip\nu\hbar}\\
&=\tau_0(q,p)+\hbar^2 \, \tau_1(q,p) + \hbar^4 \,  \tau_2(q,p)+...
\end{split}
\end{equation}
where the $\tau_n(q,p)$'s are independent of $\hbar$.
The leading term turns out just to be the local time of arrival (\ref{ltoa}).  This implies that the supraquantized TOA-operator correctly leads to the LTOA in the classical limit $\hbar \to 0$ and to the full classical time of arrival (\ref{classical}) as its extension. 

Most importantly, the leading term $\hat{\mathrm{T}}_0$ of the supraquantized TOA-operator in (\ref{supraexpand}) is exactly the Weyl-quantized TOA-operator. That is, the leading kernel $\left\langle q\left|T_0\right|q'\right\rangle$ in (\ref{expandkernel}) appears as 
\begin{equation}\label{weylltoa}
\langle q|\hat{T}_0|q'\rangle=\frac{1}{2\pi \hbar}\int_{-\infty}^{\infty} dp \,\, \tau \left(\frac{q+q'}{2},p\right)\,\mathrm{e}^{i(q-q')p/\hbar}
\end{equation}
in accordance with the Weyl quantization of the LTOA. This means that the supraquantized TOA-operator formally consists of the Weyl-quantized TOA-operator $\hat{\mathrm{T}}_0$ as the leading term plus some additional terms $\hat{\mathrm{T}}_n$ whose Weyl-Wigner transforms are dependent on even powers of $\hbar$. The additional terms $T_n(q,q')$ for $n\ge1$ in (\ref{kernelexpand}) are then interpreted as quantum corrections to the leading kernel factor $T_0(q,q')$. Likewise, the operators $\hat{\mathrm{T}}_n$ in  (\ref{supraexpand}) are the quantum corrections to the Weyl-quantized TOA-operator $\mathrm{\hat{T}}_0$. These corrections arise due to the requirement that the supraquantized TOA-operator should be canonically conjugate with the system Hamiltonian. Without explicitly solving for the $\hat{\mathrm{T}}_n$'s, it has been noted by one of us that that these corrections appear only for the case of nonlinear systems but not on linear systems \cite{Galapon2004}. 

One major issue with supraquantization is that no closed-form expression exists for each of the quantum correction $\mathrm{\hat{T}}_n$ unlike the leading term $\mathrm{\hat{T}}_0$ which is just the Weyl-quantization of the LTOA. In Ref. \cite{Galapon2004} where the supraquantized TOA-operator was introduced, the functional form of the quantum corrections were not investigated. It is believed the solution of the time kernel equation needed in the construction of the supraquantized TOA-operator is generally intractable to be useful for the case of nonlinear systems \cite{Sombillo2012}. This makes the usefulness of supraquantization limited only to linear systems and just to highlight the significance of the Weyl quantization of the classical time of arrival over other quantizations in terms of which quantization satisfies the conjugacy requirement for various quantum systems. 

In previous works, time of arrival problems involving nonlinear systems were dealt by approximating the corresponding TOA-operator as the Weyl-quantization of the classical time of arrival \cite{Galapon2012,Sombillo2014,Sombillo2018,Pablico2020} or by numerically constructing the supraquantized TOA-operator by quadrature methods \cite{Sombillo2012}. The former method is problematic since the correction terms are generally non-vanishing and important physical contents may have been missed due to the use of such approximation. On the other hand, the latter has only been done for the case of anharmonic oscillator potential and extending it to other more realistic potentials might not be trivial. 

The need for solving the quantum corrections $\hat{\mathrm{T}}_n$ in (\ref{supraexpand}) becomes significant when we consider the quantum tunneling time problem. There is still no consenus on whether the tunneling particle traverses the barrier region instantaneously or not. In Ref. \cite{Galapon2012}, one of us has found that quantum tunneling is instantaneous by taking the expectation value of the Weyl-quantized TOA-operator in the presence of a rectangular potential barrier. However, a potential barrier system is clearly a nonlinear system. Since the Weyl-quantized TOA-operator is just the leading term of the supraquantized TOA-operator, it is compelling to ask whether or not the quantum corrections provide a nonvanishing quantum tunneling time. If it will turn out that they do, it should help us determine in what time scale does quantum tunneling occur. Will it be in the attosecond regime or in the zeptosecond regime? 

Another equally important issue that highlights the importance of the quantum corrections is knowing the exact role of the time-energy canonical commutation relation to the dynamics of a legitimate TOA-operator. In Ref. \cite{Galapon2018}, it has been hypothesized that the eigenfunctions of the operator that satisfies the conjugacy relation will have the sharpest arrival at the arrival point at their respective eigenvalues. A complete supraquantized TOA-operator, with all the quantum corrections, will allow us test such hypothesis since it is the only TOA-operator in the RHS formalism that satisfies the conjugacy relation with the Hamiltonian for both linear and nonlinear systems. 

The main objective then of this paper is to undertake the explicit construction of the quantum corrections to the Weyl-quantization of the classical time of arrival, thereby completing the supraquantized TOA-operator initially introduced in Ref. \cite{Galapon2004}. Specifically, we will solve explicitly the kernel factor corrections $T_n(q,q')$ in (\ref{kernelexpand}) so that the quantum corrections $\hat{\mathrm{T}}_n$ are immediately determined in accordance with (\ref{toadef}) and (\ref{timekernel}). We will show that each kernel factor correction can be derived from the following recurrence relation
\begin{equation}\label{tnuvintro}
\begin{split}
T_n(u,v)=&\left(\frac{\mu}{2\hbar^2}\right)\sum_{r=1}^{n} \frac{1}{(2r+1)!}\frac{1}{2^{2r}}\int_{0}^{u} ds \, V^{(2r+1)}\left(\frac{s}{2}\right) \int_{0}^{v} dw \, w^{2r+1} \, T_{n-r}(s,w)\\
& \times {}_0F_1 \left(;1;\left(\frac{\mu}{2\hbar^2}\right)(v^2-w^2)\left[V \left(\frac{u}{2}\right)-V \left(\frac{s}{2}\right)\right]\right).
\end{split}
\end{equation}
for $n\ge1$ where $u=q+q'$ and $v=q-q'$. The Weyl-quantized time kernel factor $T_0(u,v)$ (\ref{tweyl}) appears as the initial condition of (\ref{tnuvintro}).

The paper is organized as follows. In Sec. \ref{sec:supraquantization}, we review the rigged Hilbert space formulation of quantum mechanics and the construction of the supraquantized TOA-operator without quantization. In Sec. \ref{sec:corrections}, we rederive the Weyl-quantized TOA-operator in terms of generating functions, a method different from the Frobenius method used in Ref. \cite{Galapon2004}. We then apply the same method in formally solving the quantum corrections and expressed them as some integrals of the interaction potential. We then investigate in detail the properties of these quantum corrections. In Sec. \ref{sec:examples}, we consider the case of quartic anharmonic oscillator potential as an example and solve the first three nonvanishing quantum corrections to the corresponding Weyl-quantized TOA-operator. We calculate the Weyl-Wigner transforms of the correction terms and show that they vanish in the classical limit $\hbar \to 0$. Finally, we conclude in Sec. \ref{sec:conclusion}.

\section{Construction of time of arrival operators without quantization}\label{sec:supraquantization}

\subsection{Time of arrival operator in the rigged Hilbert space}

In the Hilbert space formulation of quantum mechanics, a quantum mechanical system is represented by an infinite dimensional Hilbert space $\mathcal{H}$ over the complex field. A ray in $\mathcal{H}$ represents a pure state while a generally maximally symmetric densely defined operator in $\mathcal{H}$ corresponds to an observable \cite{Galapon2004,Buschm1995}. The eigenvalues of an operator provide the possible measurement values of the corresponding quantum observable. The spectrum of the operator which is the set of the eigenvalues can be discrete, continuous, or combination of both. 

It is well established that observables represented by bounded operators with discrete spectrum are defined on the whole of $\mathcal{H}$ and that their eigenvectors belong to $\mathcal{H}$. However, quantum mechanical observables are generally unbounded and that their spectrum has in general a continuous part. The eigenfunctions corresponding to contiuous spectrum are non-normalizable. To deal with these observables, one uses Dirac's bra-ket formalism which generalizes the linear algebra of Hermitian matrices for observables with discrete spectrum to observables with continuous spectrum. However, the bra-ket formalism does not make sense on Hilbert space alone but is formally and mathematically justified by the rigged Hilbert space (RHS) \cite{Madrid2005}. This is apparent when we consider non-square integrable eigenfunctions, such as $\delta(q-q_0)$ and $\mathrm{exp}(ipq/\hbar)/\sqrt{2\pi\hbar} $, which are outside of the usual Hilbert space of quantum mechanics. 

Given a Hilbert space $\mathcal{H}$, its rigged Hilbert space extension is a triplet, $\Phi^\times \supset \mathcal{H} \supset \Phi $, where $\Phi$ is a dense subspace of $\mathcal{H}$ and  $\Phi^\times$ is the space of all continuous linear functionals on $\Phi$. The subspace $\Phi $ is essentially the space of test functions while $\Phi^\times$ is the space of distributions. With the above definitions, a rigged Hilbert space can be referred to as a Hilbert space formally equipped with the theory of distributions so that singular objects become meaningful in the distributional sense.  \cite{Madrid2005}. For example, the non-square integrable eigenfunctions $\delta(q-q_0)$ and $\mathrm{exp}(ipq/\hbar)/\sqrt{2\pi\hbar} $ belong to $\Phi^\times$.

To illustrate how quantum observables appear in the rigged Hilbert space formalism, we consider a structureless particle in the real line whose corresponding system Hilbert space is $\mathcal{H}=L^2(\mathcal{R})$, the space of Lebesque square integrable functions over the real line. We choose the rigged Hilbert space $\Phi^\times \supset L^2(\mathcal{R}) \supset \Phi $, where $\Phi$ is the fundamental space of infinitely differentiable functions in the real line with compact supports and $\Phi^\times$ is the space of functionals on $\Phi$. The standard Hilbert space formulation of quantum mechanics is obtained by taking the closures in $\Phi$ with respect to the metric $L^2(\mathcal{R})$ \cite{Galapon2004,Galapon2018}. In this framework, a quantum observable $O$ appears as the mapping $\hat{O}: \Phi  \to \Phi^\times$ and is formally defined by the following integral operator
\begin{equation}
(\hat{O}\varphi)(q)=\int_{-\infty}^{\infty} \,dq' \,\langle q|\hat{O}|q'\rangle \varphi(q'),
\end{equation}
in the coordinate representation. The kernel $\langle q|\hat{O}|q'\rangle$ is required to be hermitian, $\langle q|\hat{O}|q'\rangle=\langle q'|\hat{O}|q\rangle^*$ to ensure the real valuedness of the expectation value of $\hat{O}$ in $\Phi$. The corresponding classical observable is derived by taking the limit $\hbar \to 0$ from the inverse Fourier transform
\begin{equation}\label{oinverse}
O_\hbar(q,p)=\int_{-\infty}^{\infty}d\nu \left\langle q+\frac{\nu}{2}\left|\hat{O}\right|q-\frac{\nu}{2}\right\rangle \mathrm{e}^{-i \nu p/\hbar}. 
\end{equation}
For example, the usual position $\hat{q}$ and momentum $\hat{p}$ operators appear as 
\begin{equation}\label{classicalqp}
(\hat{q}\varphi)(q)=\int_{-\infty}^{\infty} \,dq' \, q' \, \delta(q-q') \varphi(q'); \qquad (\hat{p}\varphi)(q)=\int_{-\infty}^{\infty} \,dq' \, i\hbar \,\delta'(q-q') \varphi(q'),
\end{equation}
respectively. The known classical and momentum observables are derived using (\ref{oinverse}) \cite{Galapon2018}.

Now we proceed with our current problem which is the construction of TOA-operators. The form of the local time of arrival given by (\ref{ltoa}) clearly shows the dependence of the classical time of arrival on the classical position and momentum observables. The corresponding quantum image of the classical time of arrival is then expected to be written in terms of the operators $\hat{q}$ and $\hat{p}$ in free form. But these two operators are generally represented by unbounded operators with continuous spectrum. In addition, we require that our TOA-operator to be canonically conjugate with the system Hamiltonian. But Hamiltonians, regardless of its spectrum, can also be treated within the rigged Hilbert space formalism and that the rigged Hilbert space can contain all the physically meaningful solutions of the Schr\"{o}dinger equation \cite{Madrid2002}. Adding to the fact that Dirac's bra-ket formalism is mathematically justified in the rigged Hilbert space, it is then meaningful to construct TOA-operators within the rigged Hilbert space formulation of quantum mechanics. This was exactly done in Ref. \cite{Galapon2004}, where one of us has constructed the supraquantized TOA-operator $\hat{\mathrm{T}}$ in the RHS $\Phi^\times \supset L^2(\mathcal{R}) \supset \Phi $ whose general form is defined by (\ref{toadef}). The problem is that the constructed TOA-operator in Ref.  \cite{Galapon2004} is incomplete in the sense that the quantum corrections $\hat{\mathrm{T}}_n$'s are not investigated. 

\subsection{The time kernel equation}
Supraquantization formally enters when we determine the form of the kernel  $\langle q|\hat{T}|q'\rangle$ by imposing four conditions on the supraquantized TOA-operator (\ref{toadef}). First, we require that the operator is Hermitian, i.e.,  $\hat{\mathrm{T}}=\hat{\mathrm{T}}^\dagger$,  so that it yields real expectation values. This implies that the kernel must satisfy $\langle q'|\hat{T}|q \rangle^*=	\langle q|\hat{T}|q '\rangle$. Second, the supraquantized TOA-operator must satisfy the time reversal symmetry $\Theta \hat{\mathrm{T}} \Theta ^{-1}=-\hat{\mathrm{T}}$, where $\Theta$ is the time reversal operator, so that the kernel must satisfy $\langle q|\hat{T}|q '\rangle^*=-\langle q|\hat{T}|q '\rangle$. Third, the supraquantized TOA-operator must reduce to the classical arrival time in the classical limit, i.e., $T_0(q,p)=\lim_{\hbar \to 0}\, \mathcal{T}_\hbar(q,p)$, where $\mathcal{T}_\hbar(q,p)$ is the Wigner transform of the kernel $\langle q|\hat{T}|q'\rangle$ in accordance to (\ref{oinverse}). All these properties lead to the following form of the time kernel,
\begin{equation*}
\langle q|\hat{T}|q'\rangle=\frac{\mu}{i \hbar}\, \mathrm{sgn}(q-q')\,T(q,q'),
\end{equation*}
As we noted earlier, the same form of the time kernel is derived if we use the canonical quantization of the classical time of arrival. 

The difference between the TOA-operators constructed via quantization and supraquantization lies on the explicit form of the time kernel factor $T(q,q')$ in (\ref{timekernel}). In supraquantization, the kernel factor is derived from the following canonical commutation relation  
\begin{equation}\label{commutation}
\langle \tilde{\varphi}|[\mathrm{\hat{H}}^\times,\mathrm{\hat{T}}]\varphi \rangle=i\hbar \, \langle\tilde{\varphi}|\varphi \rangle=i\hbar \, \int_{-\infty}^{\infty}dq \,  \tilde{\varphi}^*(q)\varphi(q)
\end{equation}
for all $\tilde{\varphi} \,, \varphi \, \in \, \Phi$ where $\mathrm{\hat{H}}^\times=-(\hbar^2/2\mu)\,  d^2/dq^2 + V(q)$ is the rigged Hilbert space extension of the Hamiltonian $\mathrm{\hat{H}}$ in the entire $\Phi^\times$. The above equation is to be interpreted in the distributional sense. 

Expanding the commutator in (\ref{commutation}), inserting two identity operators of the form $\mathbb{1}=\int_{-\infty}^{\infty}dq \, |q\rangle\langle q|$, and using the explicit form of the time kernel $\langle q|\hat{T}|q'\rangle$ (\ref{timekernel}),  we find
\begin{equation}\label{commutation2}
\begin{split}
\langle \tilde{\varphi}|[\mathrm{\hat{H}}^\times,\mathrm{\hat{T}}]\varphi \rangle&=\frac{\mu}{i\hbar}\int_{-\infty}^{\infty}dq\int_{-\infty}^{\infty}dq' \, T(q,q') \left[-\frac{\hbar^2}{2\mu}\left(\varphi(q')\frac{\partial^2\tilde{\varphi}^*(q)}{\partial q^2}-\tilde{\varphi}^*(q)\frac{\partial^2\varphi(q')}{\partial q^2}\right)\right]\, \mathrm{sgn}(q-q')\\
&+\frac{\mu}{i\hbar}\int_{-\infty}^{\infty}dq\int_{-\infty}^{\infty}dq' \, T(q,q') \bigg[\left(V(q)-V(q')\right)\varphi(q')\tilde{\varphi}^*(q)\bigg]\, \mathrm{sgn}(q-q').
\end{split}
\end{equation}
Performing two successive integration by parts, we arrive at
\begin{equation}\label{commutation3}
\begin{split}
\langle\tilde{\varphi}|[\mathrm{\hat{H}}^\times,\mathrm{\hat{T}}]\varphi \rangle&=i\hbar \int_{-\infty}^{\infty}dq \, \tilde{\varphi}^*(q)\varphi(q) \left(\frac{\mathrm{d}T(q,q)}{\mathrm{d}q}+\frac{\partial T(q,q')}{\partial q}+\frac{\partial T(q,q')}{\partial q'}\right)\\
&+\frac{\mu}{i\hbar}\int_{-\infty}^{\infty}dq\int_{-\infty}^{\infty}dq' \, \varphi(q')\tilde{\varphi}^*(q)\,\mathrm{sgn}(q-q')\bigg[-\frac{\hbar^2}{2\mu} \frac{\partial^2 T(q,q')}{\partial q^2}+\frac{\hbar^2}{2\mu} \frac{\partial^2 T(q,q')}{\partial q'^2}\\
&+ \left[V(q)-V(q')\right]T(q,q')\bigg].
\end{split}
\end{equation}

Note that (\ref{commutation3}) must be equal to (\ref{commutation}) for all $\tilde{\varphi}$ and $\varphi$ in $\Phi$. Since $\tilde{\varphi}$ and $\varphi$ are arbitrary and the factor $\mathrm{sgn}(q-q')$ is not identically zero, the first term of (\ref{commutation3}) should be equal to the left hand side of (\ref{commutation}) and that the second term should identically vanish. Hence, the time-energy canonical commutation relation is satisfied if and only if the time kernel factor $T(q,q')$ satisfies the following partial differential equation,
\begin{equation}\label{TKE2}
-\frac{\hbar^2}{2\mu} \frac{\partial^2 T(q,q')}{\partial q^2}+\frac{\hbar^2}{2\mu} \frac{\partial^2 T(q,q')}{\partial q'^2}+ \left[V(q)-V(q')\right]T(q,q')=0,
\end{equation}
subject to the following boundary condition
\begin{equation}\label{BC2}
\frac{\mathrm{d}T(q,q)}{\mathrm{d}q}+\frac{\partial T(q,q')}{\partial q}\bigg\vert_{q=q'}+\,\frac{\partial T(q,q')}{\partial q'}\bigg\vert_{q'=q}=1.
\end{equation}

Equation (\ref{TKE2}) is exactly the time kernel equation we referred to earlier. Of course, the boundary condition given  by (\ref{BC2}) defines a family of operators satisfying the commutation relation with the extended Hamiltonian. The specific boundary conditions are determined by requiring that the kernel factor $T(q,q')$ should lead us to the correct classical time of arrival. Also, $T(q,q')$ must be symmetric and that it should lead to the known time kernel factor for the free particle in the limit of vanishing potential, $T(q,q')=(q+q')/4.$ All of these conditions require us to set  the boundary conditions to 
\begin{equation}\label{bc}
T(q,q)=\frac{q}{2} ;      \qquad  T(q,-q)=0.
\end{equation}
The time kernel equation, coupled with the given boundary conditions, admits a unique solution for entire analytic interaction potentials \cite{Sombillo2012}.

The above results can be straightforwardly extended from the arrival point at the origin to any arbitrary arrival point $x$. For this case, the classical time of arrival is obtained by changing variables in (\ref{classical}) to $(\tilde{q}=q-x)$ so that
\begin{equation}\label{classicaltx}
T_x(\tilde{q},p)=-\mathrm{sgn}(p)\sqrt{\frac{\mu}{2}}\int_{0}^{\tilde{q}} \, \frac{d\tilde{q}'}{\sqrt{H(\tilde{q}+x,p)-V(\tilde{q}'+x)}}.
\end{equation}
Equation (\ref{classicaltx}) implies that the time of arrival problem at arbitrary arrival point $x$ is equivalent to the time of arrival problem at the origin under the potential $V(q)=V(\tilde{q}'+x)$. 

Consequently, the time kernel factor of the supraquantized TOA operator is solved from the time kernel equation subject to the potential  $V(q)=V(\tilde{q}'+x)$. For this case,  the time kernel equation given by (\ref{TKE2}) becomes
\begin{equation}\label{tkex}
-\frac{\hbar^2}{2\mu} \frac{\partial^2 T_x(\tilde{q},\tilde{q}')}{\partial \tilde{q}^2}+\frac{\hbar^2}{2\mu} \frac{\partial^2 T_x(\tilde{q},\tilde{q}')}{\partial \tilde{q}'^2}+ \left[V(\tilde{q}+x)-V(\tilde{q}'+x)\right]T_x(\tilde{q},\tilde{q}')=0,
\end{equation}
where the solution $T_x(\tilde{q},\tilde{q}')$ is subject to the boundary conditions $T_x(\tilde{q},\tilde{q})=\tilde{q}/2$ and $T_x(\tilde{q},-\tilde{q})=0$. Equation (\ref{tkex}) implies that the kernel factor $T_{x}(\tilde{q},\tilde{q}')$ is solved by a simple shift in the potential. Because of this, it is sufficient for us to consider the time of arrival problem at the origin in the developments that will folow since we only need to do an appropriate change of variables when the arrival point is different from the origin.

\subsection{Solution to the time kernel equation for entire analytic potentials}

The construction of the supraquantized TOA-operator defined by (\ref{toadef}) and (\ref{timekernel}) translates into solving the time kernel equation (\ref{TKE2}) for $T(q,q')$. Hence, we provide a brief review on the solution of the time kernel equation for arbitrary entire analytic potentials initially presented in Ref. \cite{Galapon2004}. We then show how the quantum corrections arise and determine their mathematical forms.

It will be more convenient to express the time kernel equation in canonical form. Performing a change in variable from $(q,q')$ to $(u=q+q', v=q-q')$ in (\ref{TKE2}) and (\ref{bc}), the time kernel equation assumes the form
\begin{equation}\label{tkeuv}
-\frac{2\hbar^2}{\mu} \frac{\partial^2 T(u,v)}{\partial u \partial v}+\left(V\left(\frac{u+v}{2}\right)-V\left(\frac{u-v}{2}\right)\right)T(u,v)=0,
\end{equation}
subject to the following boundary conditions
\begin{equation}\label{boundary}
T(u,0)=\frac{u}{4} ;      \qquad  T(0,v)=0.
\end{equation}

For generality, we consider an arbitrary entire analytic potential of the form given by
\begin{equation} \label{nonlinear}
V(q)=\sum_{s=1}^{\infty} a_s q^s.
\end{equation}
where $a_s$ are some coefficients. For linear systems, $a_s=0$ for $s\ge3$ while at least $a_3$ needs to be nonvanishing for nonlinear systems. The corresponding time kernel equation is found to be
\begin{equation} \label{tke2}
-2 \frac{\hbar^2}{\mu} \frac{\partial^2 T(u,v)}{\partial u \partial v} + \sum_{s=1}^{\infty} \frac{a_s}{2^{s-1}} \sum_{k=0}^{[s]} \binom{s}{2k+1}\, u^{s-2k-1} \,  v^{2k+1} \, T(u,v)=0,
\end{equation}
where $[s]=(s-1)/2$ for $s=\mathrm{odd}$ and  $[s]=s/2-1$ for $s=\mathrm{even}$. We assume an analytic solution in powers of $u$ and $v$ given by 
\begin{equation}\label{tkesol}
T(u,v) = \sum_{m,n=0}^{\infty} \alpha_{m,n} u^m v^n.
\end{equation}
for some unknown coefficients $\alpha_{m,n}$ where $\alpha_{m,n}=0$ for negative values of $m$ or $n$. The boundary conditions given by (\ref{boundary}) imply the initial conditions $\alpha_{m,0}=\delta_{m,1}/4$, and $\alpha_{0,n}=0$ for all $m$ and $n$. Substituting the assumed solution back into (\ref{tke2}) and collecting equal powers of $u$ and $v$, we find
\begin{equation}\label{coeff}
\alpha_{m,n}=\left(\frac{\mu}{2\hbar^2}\right) \frac{1}{mn}\sum_{s=1}^{\infty} \frac{a_s}{2^{s-1}}\sum_{k=0}^{[s]} \binom{s}{2k+1}\alpha_{m-s+2k,n-2k-2}.
\end{equation}

By investigating the firt few iterates of (\ref{coeff}), it can be shown that the coefficients $\alpha_{m,n}$ vanish for odd $n$ but nonvanishing for even $n$. This implies that odd powers of $v$ do not contribute in the solution given by (\ref{tkesol}). This is important as it signifies the symmetry of our solution, i.e., $T(u,v)=T(u,-v)$ or $T(q,q')=T(q',q)$ in the original coordinates, which ensures the real valuedness of the expectation value of our supraquantized TOA-operator. 

Using (\ref{coeff}) and performing some shifting of indices and reordering of summations, (\ref{tkesol}) leads to 
\begin{equation}\label{tkenonlinear}
T(u,v) = \sum_{m=0}^{\infty} \sum_{j=0}^{\infty}\, u^{m} \, v^{2j} \,	\sum_{s=0}^{j-1} \left(\frac{\mu}{2 \hbar^2}\right)^{j-s} \alpha_{m,j}^{(s)},
\end{equation}
where the new coefficients $\alpha_{m,j}^{(s)}$ satisfy the recurrence relation
\begin{equation}\label{alphasmj}
\alpha_{m,j}^{(s)}=\frac{1}{m \cdot 2j} \sum_{r=0}^{s}\sum_{l=2r+1}^{m+2r-1}\frac{a_l}{2^{l-1}}\,\binom{l}{2r+1} \, \alpha_{m-l+2r,j-r-1}^{(s-r)},
\end{equation}
for all $0 \le s \le (j-1)$. The coefficients $\alpha_{m,j}^{(s)}$ are subjected to the conditions $\alpha_{m,j}^{(0)}=\delta_{m,1}/4$ and $\alpha_{m,j}^{(s)}=0$ for $m,j \le0$. Note that the $j=0$ terms in (\ref{tkenonlinear}) are vanishing but we retained them for convenience.

Taking the Weyl-Wigner transform of (\ref{tkenonlinear}) in accordance with (\ref{oinverse}), we find 
\begin{equation}\label{wignertuv}
\mathcal{T}_\hbar(q,p)=\frac{\mu}{i\hbar}\sum_{m=0}^{\infty} \sum_{j=0}^{\infty}\, \sum_{s=0}^{j-1}2^mq^{m} \left(\frac{\mu}{2 \hbar^2}\right)^{j-s} \alpha_{m,j}^{(s)}\int_{-\infty}^{\infty}d\nu \,  \nu^{2j}\,\mathrm{sgn}(\nu)\,e^{-i\nu p\hbar}.
\end{equation}
The integral along $\nu$ is evaluated in distributional sense using the identity
\begin{equation}\label{seventh}
\int_{-\infty}^{\infty} \nu^{m-1} \mbox{sgn} (\nu) e^{-ix\nu} d\nu = \frac{2(m-1)!}{i^m x^m},
\end{equation}
(the inverse Fourier transform of \cite{Gelfand1964}, p. 360, no. 18). Equation (\ref{wignertuv}) then leads to
\begin{equation}\label{wignertuv2}
\mathcal{T}_\hbar(q,p)=\frac{\mu}{p}\sum_{m=0}^{\infty} \sum_{j=0}^{\infty}\,(-1)^j \, 2^{m+1}\, (2j)! \,\frac{q^m}{p^{2j}} \, \sum_{s=0}^{j-1} \, \left(\frac{\mu}{2}\right)^{j-s}\,\alpha_{m,j}^{(s)}\,\hbar^{2s} 
\end{equation}

It can be seen immediately from (\ref{wignertuv2}) that the $s=0$ terms (with $j>0$) are independent of $\hbar$ while the succeeding terms are dependent on  $\hbar^{2s}$ for $s\ge1$. This implies that the only contributing terms in the classical limit are the $s=0$ terms. The $s=1$ terms provide the leading $\hbar$ correction, which is of the order $\mathcal{O}(\hbar^2)$ while the succeeding terms, $s \ge 2$, provide corrections of the order $\mathcal{O}(\hbar^{2s})$.

It is then meaningful to isolate the $s=0$ terms in (\ref{tkenonlinear}) so that the solution assumes the form
\begin{equation}\label{tkesol3compact}
T(u,v) =T_{0}(u,v) +  T_{Q}(u,v),
\end{equation}
where 
\begin{equation}\label{tkesolT0gen}
T_{0}(u,v) = \sum_{m=0}^{\infty} \sum_{j=0}^{\infty} u^{m} v^{2j} 	\left(\frac{\mu}{2 \hbar^2}\right)^j \alpha_{m,j}^{(0)},
\end{equation}
\begin{equation}\label{tkesolTQ}
T_{Q}(u,v) =\sum_{m=0}^{\infty} \sum_{j=0}^{\infty} u^{m} v^{2j} 	\sum_{s=1}^{j-1} \left(\frac{\mu}{2 \hbar^2}\right)^{j-s} \alpha_{m,j}^{(s)}.
\end{equation}

It can be shown that the Weyl-Wigner transform of (\ref{tkesolT0gen}) (the $s=0$ terms in (\ref{wignertuv2})) converges to the local time of arrival which is just the expansion of the classical time of arrival at the arrival point. Hence, the leading kernel factor $T_0(u,v)$ can be closed in integral form given by
\begin{equation}\label{t0uv}
T_{0}(u,v)=\frac{1}{4}\int_{0}^{u} ds \, {}_0F_{1}\left(;1;\left(\frac{\mu}{2\hbar^2}\right)v^2\left[V\left(\frac{u}{2}\right)-V\left(\frac{s}{2}\right)\right]\right).
\end{equation}
in accordance with (\ref{weylltoa}).
Recall that $T_0(u,v)$ is just equal to the Weyl-quantized time kernel factor in the orginal $(q,q')$ coordinates given by (\ref{tweyl}), i.e., $T_0(q,q')=T_W(q,q')$. The term $T_Q(u,v)$ defined by (\ref{tkesolTQ}) consists all the kenerl factor corrections to the leading term  $T_{0}(u,v)$. 

To simplify the corrections $T_Q(u,v)$, we rewrite (\ref{tkesolTQ}) by series rearrangement to get
\begin{equation}\label{tkesolTQ2}
T_{Q}(u,v) =\sum_{n=1}^{\infty}T_n(u,v),
\end{equation}
where
\begin{equation}\label{tnuv}
T_n(u,v)=\sum_{m=0}^{\infty}\sum_{j=0}^{\infty}u^m v^{2j+2n+2}\left(\frac{\mu}{2 \hbar^2}\right)^{j+1}\alpha^{(n)}_{m,j+n+1}.
\end{equation}
Since the kernel factor $T_n(u,v)$ corresponds to the contribution $\mathcal{O}(\hbar^{2n})$ in the classical limit, we formally refer it as the $n$th kernel factor correction.
The appropriate boundary conditions for the corrections $T_n(u,v)$ can be easily determined by noting that the leading term $T_0(u,v)$ itself satisfies the full boundary conditions given by (\ref{boundary}), i.e., $T_0(u,0)=u/4$ and $T_0(0,v)=0$. This means that the correction terms $T_Q(u,v)$ satisfy the condition $T_{Q}(u,0)=T_{Q}(0,v)=0$. Hence, each kernel factor correction satisfies
\begin{equation}\label{bccorrect}
T_{n}(u,0)=T_{n}(0,v)=0
\end{equation} 

With the form of $T_Q(u,v)$, the supraquantized TOA-operator formally appears as the expansion $\hat{\mathrm{T}}=\hat{\mathrm{T}}_0+\hat{\mathrm{T}}_1+\hat{\mathrm{T}}_2+...$. The leading term $\hat{\mathrm{T}}_0$ is the Weyl-quantized TOA-operator and the $\hat{\mathrm{T}}_n$'s for $n\ge1$ are the quantum corrections to the Weyl quantization of the classical time of arrival. It will be shown in a clear manner later that the appearance of the quantum corrections suggests the failure of Weyl quantization in consistently satisfying the required commutator values with the Hamiltonian. 

\section{The quantum corrections and the complete solution of the time kernel equation}\label{sec:corrections}

In Ref. \cite{Galapon2004}, no analysis was done on the functional structure of the quantum corrections. In this paper, we will explicitly show that that the kernel factor corrections can also be expressed as integrals dependent on the interaction potential, similar to the leading kernel $T_0(u,v)$. 

We note that $T_0(u,v)$ (\ref{t0uv}) was derived in Ref. \cite{Galapon2004} by comparing its Weyl-Wigner transform to the local time of arrival given by (\ref{ltoa}). Hence, the coefficients $\alpha_{m,j}^{(0)}$ in (\ref{tkesolT0gen}) were not explicitly solved. However, the same method cannot be extended to the quantum corrections since they vanish in the classical limit. For example, we want to determine the leading kernel factor correction $T_1(u,v)$ which consists of the $s=1$ terms in (\ref{tkesolTQ}), that is,
\begin{equation}\label{t1uv}
T_1(u,v)= \sum_{m=0}^{\infty} \sum_{j=0}^{\infty} u^{m} v^{2j} 	\left(\frac{\mu}{2 \hbar^2}\right)^{j-1} \alpha_{m,j}^{(1)}
\end{equation}
Using (\ref{alphasmj}), the coefficients $\alpha_{m,j}^{(1)}$ are determined by solving the following recurrence relation
\begin{equation} \label{alphamjsa}
\alpha_{m,j}^{(1)}=\frac{1}{m \cdot 2j}\sum_{l=1}^{m-1} \frac{l a_l}{2^{l-1}}\alpha_{m-l,j-1}^{(1)} + \frac{1}{m \cdot 2j} \sum_{l=3}^{m+1} \frac{a_l}{2^{l-1}}\binom{l}{3}\alpha_{m-l+2,j-2}^{(0)}.
\end{equation}
The above equation cannot be solved without first determining the coefficients $\alpha_{m,j}^{(0)}$ of the leading kernel factor $T_{0}(u,v)$. But recall that these coefficients are also unknown. One can try solving ($\ref{alphamjsa}$) iteratively by assuming few values of $m$ and $j$ and then infer a general form of the coefficients $\alpha_{m,j}^{(1)}$. Afterwhich, one can formally prove the result via mathematical induction. However, such method is too cumbersome and impractical as we consider the higher order correction terms. 

What we want then is to determine the $T_n(u,v)$'s without explicitly solving for the coefficients $\alpha_{m,j}^{(n)}$ in (\ref{tnuv}). This can be done by noting that the kernel corrections $T_n(u,v)$ are, in fact, generating functions of the coefficients $\alpha_{m,j}^{(n)}$. Hence, we should be able to derive a closed-form expression for each of the kernel factor $T_n(u,v)$ using the known techniques of obtaining generating functions, such as the series rearrangement technique \cite{McBride1971,Srivastava1984a}. 

\subsection{Alternative derivation of the leading term}\label{subsec:alternative}

Before we proceed with the quantum corrections, we rederive the leading time kernel $T_{0}(u,v)$ by series rearrangement technique followed by the method of successive approximations. This is done to show that interpreting the kernel factor corrections $T_n(u,v)$ as generating functions to obtain their integral forms is more staightforward than using the Frobenius method originally prescribed in Ref. \cite{Galapon2004}.

We go back to (\ref{tkesolT0gen}). Isolating the $j=0$ term and using the initial condition $\alpha_{m,0}^{(0)}=\delta_{m,1}/4$, we get
\begin{equation}\label{tkesolT0genb}
T_0(u,v) = \frac{u}{4} + \sum_{m=0}^{\infty} \sum_{j=1}^{\infty} u^{m} v^{2j} 	\left(\frac{\mu}{2 \hbar^2}\right)^j \alpha_{m,j}^{(0)},
\end{equation}
where the coefficients $\alpha_{m,j}^{0}$ satisfy the following recurrence relation
\begin{equation}\label{alphas_00}
\alpha_{m,j}^{(0)}=\frac{1}{m \cdot 2j} \sum_{s=1}^{m-1} \frac{s a_s}{2^{s-1}}\alpha_{m-s,j-1}^{(0)}.
\end{equation}
Direct substitution of (\ref{alphas_00}) into (\ref{tkesolT0genb}), we have
\begin{equation}\label{tkesolT0gend}
T_{0}(u,v)=\frac{u}{4}+\sum_{m=0}^{\infty}\sum_{j=1}^{\infty}\frac{u^mv^{2j}}{m \cdot 2j}\left(\frac{\mu}{2\hbar^2}\right)^j  \sum_{s=1}^{m-1} \frac{s a_s}{2^{s-1}}\alpha_{m-s,j-1}^{(0)}.
\end{equation}
Shifting indices from $l$ to $l-1$, $j$ to $j-1$, and $m$ to $m-2$, we arrive at
\begin{equation}\label{tkesolT0gene}
T_{0}(u,v)=\frac{u}{4}+\sum_{m=0}^{\infty}\sum_{j=0}^{\infty} \frac{u^{m+2}}{m+2} \frac{v^{2j+2}}{2j+2} \left(\frac{\mu}{2\hbar^2}\right)^{j+1}\sum_{l=0}^{m} \frac{(l+1)a_{l+1}}{2^l} \alpha^{(0)}_{m-l+1,j}.
\end{equation}

Notice that $T_0(u,v)$ involves three summations along $m$, $j$, and $l$. Our immediate task is to close at least one of the three sums. This can be facilitated by using the following elementary integrals,
\begin{equation}\label{integral1}
\int_{0}^{u} ds \, s^{m+1}=\frac{u^{m+2}}{m+2}, \qquad \int_{0}^{v} dw \, w^{2j+1}=\frac{v^{2j+2}}{2j+2},
\end{equation}
so that (\ref{tkesolT0gene}) can be rewritten as 
\begin{equation}\label{tkesolT0genf}
T_{0}(u,v)=\frac{u}{4}+\sum_{m=0}^{\infty}\sum_{j=0}^{\infty} \int_{0}^{u} ds \, s^{m+1}\int_{0}^{v} dw \, w^{2j+1} \left(\frac{\mu}{2\hbar^2}\right)^{j+1}\sum_{l=0}^{m} \frac{(l+1)a_{l+1}}{2^l} \alpha^{(0)}_{m-l+1,j}.
\end{equation}

Because of the convergence of the integrals, we can safely interchange the order of summations and integrations leading to 
\begin{equation}\label{tkesolT0geng}
T_{0}(u,v)=\frac{u}{4}+\int_{0}^{u} ds \, s\int_{0}^{v} dw \, w\sum_{m=0}^{\infty}\sum_{j=0}^{\infty}\sum_{l=0}^{m}s^m \, w^{2j}\, \left(\frac{\mu}{2\hbar^2}\right)^{j+1} \frac{(l+1)a_{l+1}}{2^l} \alpha^{(0)}_{m-l+1,j}.
\end{equation}
Using the following summation identity \cite{Srivastava1984a},
\begin{equation}\label{identitysum}
\sum_{n=0}^{\infty}\sum_{k=0}^{n}B(k,n)=\sum_{n=0}^{\infty}\sum_{k=0}^{\infty}B(k,n+k), 
\end{equation}
Equation (\ref{tkesolT0geng}) can be rewritten as 
\begin{equation}\label{tkesolT0genh}
T_{0}(u,v)=\frac{u}{4}+ \int_{0}^{u} ds \, s \sum_{l=0}^{\infty}(l+1)a_{l+1}\left(\frac{s}{2}\right)^l\int_{0}^{v} dw \,w \sum_{m=0}^{\infty}\sum_{j=0}^{\infty}s^m \, w^{2j}\, \left(\frac{\mu}{2\hbar^2}\right)^{j+1}  \alpha^{(0)}_{m+1,j}.
\end{equation}
Shifting indices from $l+1$ to $l $, and $m+1$ to $m$ leads to 
\begin{equation}\label{tkesolT0genhh}
T_{0}(u,v)=\frac{u}{4}+ \int_{0}^{u} ds \, \sum_{l=1}^{\infty}l \, a_{l}\left(\frac{s}{2}\right)^{l-1}\int_{0}^{v} dw \,w \sum_{m=0}^{\infty}\sum_{j=0}^{\infty}s^m \, w^{2j}\, \left(\frac{\mu}{2\hbar^2}\right)^{j+1}  \alpha^{(0)}_{m,j}.
\end{equation}

We can now close the infinite series along $l$ by noting that our potential given by (\ref{nonlinear}) implies the following partial derivative
\begin{equation}\label{potderivb}
V'\left(\frac{s}{2}\right)=\sum_{l=1}^{\infty}l \,a_l \, \left(\frac{s}{2}\right)^{l-1},
\end{equation}
so that (\ref{tkesolT0genhh}) can be simplified as 
\begin{equation}\label{tkesolT0geni}
T_{0}(u,v)=\frac{u}{4}+ \left(\frac{\mu}{2\hbar^2}\right)\int_{0}^{u} ds \,  V'\left(\frac{s}{2}\right)\int_{0}^{v} dw \,w \sum_{m=0}^{\infty}\sum_{j=0}^{\infty}s^m \, w^{2j}\, \left(\frac{\mu}{2\hbar^2}\right)^{j}  \alpha^{(0)}_{m,j}.
\end{equation}

Notice that the infinite series inside the integral along $w$ is just $T_{0}(u,v)$ defined in (\ref{tkesolT0gen}) but in $(s,v) $ coordinates, i.e.,  $T_{0}(u,v) \to T_{0}(s,w)$. Hence, (\ref{tkesolT0geni}) simplifies to the following integral equation, 
\begin{equation}\label{tkesolT0genk}
T_{0}(u,v)=\frac{u}{4}+ \left(\frac{\mu}{2\hbar^2}\right)\int_{0}^{u} ds \,  V'\left(\frac{s}{2}\right)\int_{0}^{v} dw \,w \, T_0(s,w).
\end{equation}
We immediately see from (\ref{tkesolT0genk}) that the boundary conditions, $T_0(u,0)=u/4$, and $T_0(0,v)=0$, are satisfied. It is straightforward to convert (\ref{tkesolT0genk}) as a partial differential equation. Taking $\partial^2/\partial v\partial u$ and using the Leibniz integral rule given by
\begin{equation}\label{leibniz}
\frac{d}{dx}\left(\int_{a}^{x}f(x,t) \, dt\right)=f(x,x)+\int_{a}^{x} \frac{\partial f(x,t)}{\partial x} \, dt,
\end{equation}
we immediately find the following partial differential equation for $T_0(u,v)$
\begin{equation}\label{t0uvpde}
\frac{\partial T_0(u,v)}{\partial v \partial u}= \left(\frac{\mu}{2\hbar}\right)V'\left(\frac{u}{2}\right)\, v \, T_0(u,v).
\end{equation}
We will show later that the kernel factor corrections $T_n(u,v)$ are always vanishing for linear systems so that (\ref{t0uvpde}) is already the time kernel equation for linear systems. Likewise, the existence and uniqueness of the solution $T_0(u,v)$ is already guaranteed for this case. 

An advantage of our method here, compared to the Frobenius method used in Ref. \cite{Galapon2004}, is that we are able to derive a partial differential equation (\ref{t0uvpde}) satisfied by the leading kernel $T_0(u,v)$ alone, applicable for both linear and nonlinear systems, without solving explicitly for the coefficients $\alpha^{(0)}_{m,j}$ in (\ref{alphas_00}). This differential equation, coupled with the boundary conditions, can be used to validate any result for $T_0(u,v)$ given a specific potential. In addition, it will become important later when we show that the full solution $T(u,v)$ satisfies the time kernel equation (\ref{tkeuv}) for arbitrary potentials. 

We now go back to the integral equation involving $T_0(u,v)$ given by (\ref{tkesolT0genk}). A quick look should already allow us to see the dependence of the leading kernel factor $T_0(u,v)$ to the potential $V(q)$. The problem, however, is the appearance of $T_0(s,w)$ on the other side of the equation. We then solve the integral equation in (\ref{tkesolT0genk}) using the method of successive approximations, also known as the Picard iteration. This is done by noting that the first term of (\ref{tkesolT0genk}) is already a solution so that when either $u=0$ or $v=0$, the necessary boundary conditions readily emerged. We then choose 
\begin{equation}\label{t000uv}
T_{0,0}(u,v)=\frac{u}{4}
\end{equation}
as our zeroth order approximation of  (\ref{tkesolT0genk}). The $n$th order approximation for $n \ge 1$ is determined from the following recurrence relation,
\begin{equation} \label{iterationeq}
T_{0,n}(u,v)=\frac{u}{4}+ \left(\frac{\mu}{2\hbar^2}\right)\int_{0}^{u} ds \,  V'\left(\frac{s}{2}\right)\int_{0}^{v} dw \,w \, T_{0,n-1}(s,w).
\end{equation}
Equation (\ref{tkesolT0genk}) is retrieved from (\ref{iterationeq}) in the limit $n \to \infty$ so that the solution $T_0(u,v)$ is determined from the limit
$T_0(u,v)= \lim_{n\to \infty} T_{0,n}(u,v).$ 

Given (\ref{iterationeq}), we find the first few approximations of (\ref{tkesolT0genk}). For $n=1$, (\ref{iterationeq}) leads to 
\begin{equation}\label{t01uv}
\begin{split}
T_{0,1}(u,v)&=\frac{u}{4}+ \left(\frac{\mu}{2\hbar^2}\right)\int_{0}^{u} ds \,  V'\left(\frac{s}{2}\right)\int_{0}^{v} dw \,w \, T_0^{(0)}(s,w)\\
&=\frac{u}{4}+ \left(\frac{\mu}{2\hbar^2}\right)\frac{v^2}{8}\int_{0}^{u} ds \, s \, V'\left(\frac{s}{2}\right).
\end{split}
\end{equation}
We perform an integration by parts to arrive at the following first order approximation of $T_0(u,v)$
\begin{equation}\label{t01uvb}
T_{0,1}(u,v)=\frac{u}{4}+ \left(\frac{\mu}{2\hbar^2}\right)\frac{v^2}{4}\int_{0}^{u} ds \, \left[V\left(\frac{u}{2}\right)-V\left(\frac{s}{2}\right)\right].
\end{equation}

On the other hand, the second order approximation is determined by setting $n=2$ in (\ref{iterationeq}) leading to
\begin{equation}\label{t02uv}
\begin{split}
T_{0,2}(u,v)&=\frac{u}{4}+ \left(\frac{\mu}{2\hbar^2}\right)\int_{0}^{u} ds \,  V'\left(\frac{s}{2}\right)\int_{0}^{v} dw \,w \, T_0^{(1)}(s,w)\\
&=T_{0,1}(u,v)+ \left(\frac{\mu}{2\hbar^2}\right)^2 \frac{v^4}{16}\int_{0}^{u} ds \,  V'\left(\frac{s}{2}\right)\int_{0}^{s} ds' \, \left[V\left(\frac{s}{2}\right)-V\left(\frac{s'}{2}\right)\right].
\end{split}
\end{equation}
We rewrite (\ref{t02uv}) by exploiting the following equality
\begin{equation}\label{derivative}
\frac{\partial }{\partial s}\left[V\left(\frac{s}{2}\right)-V\left(\frac{s'}{2}\right)\right]^2 =V'\left(\frac{s}{2}\right)\left[ V\left(\frac{s}{2}\right)-V\left(\frac{s'}{2}\right)\right],
\end{equation}
and using the Leibniz integral rule given by (\ref{leibniz}) so that (\ref{t02uv}) becomes
\begin{equation}\label{t02uvd}
T_{0,2}(u,v)=T_{0,1}(u,v)+ \left(\frac{\mu}{2\hbar^2}\right)^2 \frac{v^4}{16}\int_{0}^{u} ds \,  \left[V\left(\frac{u}{2}\right)-V\left(\frac{s}{2}\right)\right]^2.
\end{equation}
Inserting $T_{0,1}(u,v)$ into (\ref{t02uvd}), we finally find the following second order approximation of $T_0(u,v)$
\begin{equation} \label{t02uvc}
\begin{split}
T_{0,2}(u,v)=\frac{u}{4}+ \left(\frac{\mu}{2\hbar^2}\right)\frac{v^2}{4}\int_{0}^{u} ds \, \left[V\left(\frac{u}{2}\right)-V\left(\frac{s}{2}\right)\right]+ \left(\frac{\mu}{2\hbar^2}\right)^2 \frac{v^4}{16}\int_{0}^{u} ds \,  \left[V\left(\frac{u}{2}\right)-V\left(\frac{s}{2}\right)\right]^2.
\end{split}
\end{equation}
Following similar steps, the third order approximation, which is the $n=3$ case of (\ref{iterationeq}), leads to 
\begin{equation}\label{t03uv}
\begin{split}
T_{0,3}(u,v)=&\frac{u}{4}+ \left(\frac{\mu}{2\hbar^2}\right)\frac{v^2}{4}\int_{0}^{u} ds \, \left[V\left(\frac{u}{2}\right)-V\left(\frac{s}{2}\right)\right] \left(\frac{\mu}{2\hbar^2}\right)^2 \frac{v^4}{16}\int_{0}^{u} ds \,  \left[V\left(\frac{u}{2}\right)-V\left(\frac{s}{2}\right)\right]^2\\
&+ \left(\frac{\mu}{2\hbar^2}\right)^2 \frac{v^6}{144}\int_{0}^{u} ds \,  \left[V\left(\frac{u}{2}\right)-V\left(\frac{s}{2}\right)\right]^3.
\end{split}
\end{equation}

The forms of the first three aproximations imply the following general form for arbitrary $n$,
\begin{equation}\label{t0nuv}
T_{0,n}(u,v)=\frac{1}{4}\sum_{k=0}^{n}\left(\frac{\mu}{2\hbar^2}\right)^k \frac{v^{2k}}{(1)_k k!}\int_{0}^{u} ds \,\left[V\left(\frac{u}{2}\right)-V\left(\frac{s}{2}\right)\right]^k.
\end{equation}
Equation (\ref{t0nuv}) is formally proven via mathematical induction. Assuming that the above equation is valid for some $n=k \ge 0$, the next iterate which is $n=k+1$ also holds true, validating (\ref{t0nuv}).

Because of the continuity of the potential and convergence of the integral, we can interchange the order of summation and integration. Taking the limit $n \to \infty$, we find 
\begin{equation}\label{t0nuvb}
T_0(u,v)=\frac{1}{4}\int_{0}^{u} ds \,\sum_{k=0}^{\infty}\left(\frac{\mu}{2\hbar^2}\right)^k \frac{v^{2k}}{(1)_k k!}\left[V\left(\frac{u}{2}\right)-V\left(\frac{s}{2}\right)\right]^k.
\end{equation}
The infinite series along $k$ can be simplified by using the definition of the hypergeometric function given by
\begin{equation}\label{hypergeom}
{}_0F_{1}(;b;z)=\sum_{k=0}^{\infty}\frac{z^k}{(b)_k k!}.
\end{equation}
Hence, we finally arrive at the following solution
\begin{equation}\label{t0uvfin}
T_0(u,v)=\frac{1}{4}\int_{0}^{u} ds \, {}_0F_{1}\left(;1;\left(\frac{\mu}{2\hbar^2}\right)v^2\left[V\left(\frac{u}{2}\right)-V\left(\frac{s}{2}\right)\right]\right),
\end{equation}
which is exactly the same result with (\ref{t0uv}). It is straightforward to show that $T_0(u,v)$ satisfies the boundary conditions $T_0(u,0)=u/4$ and $T_0(0,v)=0$.  It is also symmetric, $T_0(u,v)=T_0(u,-v)$ so that the expectation value of the corresponding operator, the Weyl-quantized TOA-operator  $\mathrm{\hat{T}}_0$, is real valued. 

Note that we did not solve for the coefficients $\alpha_{m,j}^{(0)}$ in (\ref{tkesolT0gen}) but we are still able to derive (\ref{t0uvfin}), the time kernel factor of the Weyl-quantized TOA-operator. We then extend our method here into the derivation of the kernel factor corrections $T_n(u,v)$ for $n\ge 1$ and expressed them as some integrals of the interaction potential.

\subsection{The leading quantum correction}\label{subsec:leading}

We now go back to the leading quantum correction $T_1(u,v)$ we introduced earlier in (\ref{t1uv})  where the coefficients $\alpha_{m,j}^{(1)}$ satisfy the recurrence relation given by (\ref{alphamjsa}). Direct substitution of (\ref{alphamjsa}) into (\ref{t1uv}) leads to
\begin{equation}\label{alphamj1b}
\begin{split}
T_1(u,v)&=\sum_{m=0}^{\infty}\sum_{j=0}^{\infty}u^m v^{2j}\left(\frac{\mu}{2\hbar^2}\right)^{j-1}\frac{1}{m \cdot 2j}\sum_{l=1}^{m-1} \frac{l a_l}{2^{l-1}}\alpha_{m-l,j-1}^{(1)}\\ &+\sum_{m=0}^{\infty}\sum_{j=0}^{\infty}u^m v^{2j}\left(\frac{\mu}{2\hbar^2}\right)^{j-1} \frac{1}{m \cdot 2j} \sum_{l=3}^{m+1} \frac{a_l}{2^{l-1}}\binom{l}{3}\alpha_{m-l+2,j-2}^{(0)}.
\end{split}
\end{equation}
Performing series rearrangements, shifting of indices, and using the following relations
\begin{equation}
\int_{0}^{u} ds \, s^{m+l+1}=\frac{u^{m+l+2}}{m+l+2}, \qquad \int_{0}^{v} dw \, w^{2j+1}=\frac{v^{2j+2}}{2j+2},
\end{equation}
Equation (\ref{alphamj1b}) simplifies to
\begin{equation}\label{t1uvinteg}
\begin{split}
T_1(u,v)=&\left(\frac{\mu}{2\hbar^2}\right)\int_{0}^{u} ds \, V'\left(\frac{s}{2}\right) \int_{0}^{v} dw \, w \, T_1(s,w)+\frac{1}{24}\left(\frac{\mu}{2\hbar^2}\right)\int_{0}^{u} ds \, V'''\left(\frac{s}{2}\right) \int_{0}^{v} dw \, w^3 \, T_0(s,w).
\end{split}
\end{equation}

Taking $\partial^2/\partial u\partial v$ and using again the Leibniz integral rule given by (\ref{leibniz}), we get the following partial differential equation for $T_1(u,v)$
\begin{equation}\label{t1uvpde}
\frac{\partial^2 T_1(u,v)}{\partial v \partial u}=\left(\frac{\mu}{2\hbar^2}\right)V'\left(\frac{u}{2}\right)v\,T_1(u,v)+\frac{1}{24}\left(\frac{\mu}{2\hbar^2}\right) v^3 \, V'''\left(\frac{u}{2}\right) T_0(u,v).
\end{equation}
It is straightforward to show the uniqueness of the solution $T_1(u,v)$ with boundary conditions $T_1(u,0)=T_1(0,v)=0$. Suppose that $T_{1,a}(u,v)$ and $T_{1,b}(u,v)$ both satisfy (\ref{t1uvpde}). Since the leading kernel factor $T_0(u,v)$ is unique, it can be shown using the triangle inequality that $|T_{1,a}(u,v)$ - $T_{1,b}(u,v)|\to 0$ so that $T_{1,a}(u,v)$ = $T_{1,b}(u,v)$. Hence, $T_1(u,v)$ is also unique. In fact, the uniqueness of $T_1(u,v)$ is also guaranteed by the use of the method of succesive approximations later.

Notice that (\ref{t1uvpde}) is dependent on $T_1(s,w)$ and $T_0(s,w)$ but the latter is just the leading kernel factor which is already known at this point. To solve for $T_1(u,v)$, we apply again the method of succesive approximations used in Sec. (\ref{subsec:alternative}). Since we are solving for $T_1(u,v)$, our zeroth order approximation is the second term of (\ref{t1uvinteg}), that is,
\begin{equation}\label{t10uv}
T_{1,0}(u,v)=\frac{1}{24}\left(\frac{\mu}{2\hbar^2}\right)\int_{0}^{u} ds \, V'''\left(\frac{s}{2}\right) \int_{0}^{v} dw \, w^3 \, T_0(s,w).
\end{equation}
The $n$th order approximation can theb be determined from the following equation,
\begin{equation}\label{t1nuv}
T_{1,n}(u,v)=T_{1,0}(u,v)+\left(\frac{\mu}{2\hbar^2}\right)\int_{0}^{u} ds \, V'\left(\frac{s}{2}\right) \int_{0}^{v} dw \, w \, T_{1,n-1}(s,w).
\end{equation}
The solution $T_1(u,v)$ of the integral equation in (\ref{t1uvinteg}) is derived by taking the limit, $T_1(u,v)=\lim_{n\to \infty}T_{1,n}(u,v)$.

From (\ref{t1nuv}), we determine the first few iterates and also infer the general form for arbitrary $n$. For $n=1$, we have
\begin{equation}\label{t10uv-1}
\begin{split}
T_{1,1}(u,v)&=\frac{1}{24}\left(\frac{\mu}{2\hbar^2}\right)\int_{0}^{u} ds \, V'''\left(\frac{s}{2}\right) \int_{0}^{v} dw \, w^3 \, T_0(s,w)\\
&+\frac{1}{24}\left(\frac{\mu}{2\hbar^2}\right)^2\int_{0}^{u} ds \, V'''\left(\frac{s}{2}\right)\left[V \left(\frac{u}{2}\right)-V \left(\frac{s}{2}\right)\right] \int_{0}^{v} dw \, w^3 (v^2-w^2)\, T_0(s,w).
\end{split}
\end{equation}
For $n=2$, we have
\begin{equation}\label{t10uv-2}
\begin{split}
T_{1,2}(u,v)&=\frac{1}{24}\left(\frac{\mu}{2\hbar^2}\right)\int_{0}^{u} ds \, V'''\left(\frac{s}{2}\right) \int_{0}^{v} dw \, w^3 \, T_0(s,w)\\
&+\frac{1}{24}\left(\frac{\mu}{2\hbar^2}\right)^2\int_{0}^{u} ds \, V'''\left(\frac{s}{2}\right)\left[V \left(\frac{u}{2}\right)-V \left(\frac{s}{2}\right)\right] \int_{0}^{v} dw \, w^3 (v^2-w^2)\, T_0(s,w)\\
&+\frac{1}{96}\left(\frac{\mu}{2\hbar^2}\right)^3\int_{0}^{u} ds V'''\left(\frac{s}{2}\right)\left[V \left(\frac{u}{2}\right)-V \left(\frac{s}{2}\right)\right]^2 \int_{0}^{v} dw \, w^3 (v^2-w^2)^2\, T_0(s,w).
\end{split}
\end{equation}
Doing the same calculations for $n \ge 3$, we infer the following form of $T_{1,n}(u,v)$ for abitrary $n$
\begin{equation}\label{t10uvn}
\begin{split}
T_{1,n}(u,v)=&\left(\frac{\mu}{48\hbar^2}\right)\int_{0}^{u} ds \, V'''\left(\frac{s}{2}\right) \int_{0}^{v} dw \, w^3 \, T_0(s,w) \\
&\times \sum_{k=0}^{n}\frac{1}{(1)_k k!}\left(\frac{\mu}{2\hbar^2}\right)^k(v^2-w^2)^k \left[V \left(\frac{u}{2}\right)-V \left(\frac{s}{2}\right)\right]^k.
\end{split}
\end{equation}

Equation (\ref{t10uvn}) is also proven formally via mathematical induction. Taking the limit $n \to \infty$ and using the definition of the hypergeometric function given by (\ref{hypergeom}), we find the leading kernel factor correction to be
\begin{equation}\label{t10uvfinb}
\begin{split}
T_1(u,v)&=\left(\frac{\mu}{48\hbar^2}\right)\int_{0}^{u} ds \, V'''\left(\frac{s}{2}\right) \int_{0}^{v} dw \, w^3 \, T_0(s,w)\,\,{}_0F_1 \left(;1;\left(\frac{\mu}{2\hbar^2}\right)(v^2-w^2)\left[V \left(\frac{u}{2}\right)-V \left(\frac{s}{2}\right)\right]\right)
\end{split}
\end{equation}
in its integral form. Equation (\ref{t10uvfinb}) clearly shows the dependence of the leading kernel correction $T_1(u,v)$ on the potential $V(q)$ and the leading kernel factor $T_0(u,v)$ which is also dependent on the potential. It is straightforward to show that (\ref{t10uvfinb}) satisfies the partial differential equation for the leading correction given by (\ref{t1uvpde}) subject to the boundary conditions $T_1(u,0)=T_1(0,v)=0$. 

Notice that if we consider linear systems of the form $V(q)=a +b q + cq^2$, the leading correction $T_1(u,v)$ immediately vanishes since $V'''(s/2)=0$. For nonlinear systems, the leading correction has a non-zero contribution. This vanishing and non-vanishing property of $T_1(u,v)$ for linear and non-linear systems, respectively, is actually a general property of all the quantum corrections $T_n(u,v)$ as we will show in the next subsection. 

\subsection{General expression for the nth order quantum correction}\label{subsec:nthorder}

We now determine a general expression for the $n$th order kernel factor correction $T_n(u,v)$. We show that it can also be expressed as some integral of the potential using the same methodology discussed in the previous subsections. Using (\ref{alphasmj}) and (\ref{tnuv}), the explicit form of $T_n(u,v)$ is given by
\begin{equation}\label{tnuvapp}
	T_n(u,v)=\sum_{r=0}^{n}\sum_{m=0}^{\infty}\sum_{j=0}^{\infty}\sum_{l=2r+1}^{m+2r-1} \, \frac{u^m v^{2j+2n+2}}{m \cdot 2(j+n+1)} \left(\frac{\mu}{2\hbar^2}\right)^{j+1} \frac{a_l}{2^{l-1}} \binom{l}{2r+1} \alpha^{(n-r)}_{m-l+2r,j+n-r}.
\end{equation} 
Taking advantage of the following elementary integrals
\begin{equation}
\int_{0}^{u} ds\, s^{m-1}=\frac{u^m}{m}, \qquad \int_{0}^{v}dw \, w^{2j+2n+1}=\frac{v^{2j+2n+2}}{2j+2n+2},
\end{equation}
Equation (\ref{tnuvapp}) can be rewritten as
\begin{equation}\label{tnuvapp2}
\begin{split}
T_n(u,v)=&\sum_{r=0}^{n}\sum_{m=0}^{\infty}\sum_{j=0}^{\infty}\sum_{l=2r+1}^{m+2r-1} \, \int_{0}^{u} ds\, s^{m-1} \int_{0}^{v}dw \, w^{2j+2n+1}  \left(\frac{\mu}{2\hbar^2}\right)^{j+1} \frac{a_l}{2^{l-1}}\binom{l}{2r+1} \alpha^{(n-r)}_{m-l+2r,j+n-r}.
\end{split}
\end{equation} 

The convergence of the above equation allows us to safely interchange the order of integrations and summations. Doing so leads to
\begin{equation}\label{tnuvapp3}
\begin{split}
T_n(u,v)=&\sum_{r=0}^{n}\,\int_{0}^{u} ds\, s^{-1} \int_{0}^{v}dw \, w \sum_{m=0}^{\infty}\sum_{j=0}^{\infty}\sum_{l=2r+1}^{m+2r-1} \, s^m \, w^{2j+2n} \left(\frac{\mu}{2\hbar^2}\right)^{j+1} \frac{a_l}{2^{l-1}} \binom{l}{2r+1} \alpha^{(n-r)}_{m-l+2r,j+n-r}.
\end{split}
\end{equation} 
We perform a shift in indices from $l$ to $l+2r+1$ and $m$ to $m+2$ to find
\begin{equation}\label{tnuvapp4}
\begin{split}
T_n(u,v)=&\sum_{r=0}^{n}\,\int_{0}^{u} ds\, s \int_{0}^{v}dw \, w \sum_{m=0}^{\infty}\sum_{j=0}^{\infty}\sum_{l=0}^{m} \, s^m \, w^{2j+2n} \left(\frac{\mu}{2\hbar^2}\right)^{j+1} \frac{a_{l+2r+1}}{2^{l+2r}} \\
&\times \binom{l+2r+1}{2r+1} \alpha^{(n-r)}_{m-l+1,j+n-r}.
\end{split}
\end{equation} 

We want to simplify (\ref{tnuvapp4}) by decoupling the sum along $l$ from the other infinite series. This is facilitated by using the summation identity given by (\ref{identitysum}) so that we find
\begin{equation}\label{tnuvapp5}
\begin{split}
T_n(u,v)=&\sum_{r=0}^{n}\,\int_{0}^{u} ds\, s \int_{0}^{v}dw \, w \,\sum_{l=0}^{\infty} \, \binom{l+2r+1}{2r+1}\,\frac{s^l}{2^{l+2r}} \, \,a_{l+2r+1}\\
&\times \sum_{m=0}^{\infty}\sum_{j=0}^{\infty} s^m \, w^{2j+2n} \left(\frac{\mu}{2\hbar^2}\right)^{j+1} \alpha^{(n-r)}_{m+1,j+n-r}.
\end{split}
\end{equation} 
Shifting index from $l$ to $l-2r-1$ allows us to write
\begin{equation}\label{tnuvapp6}
\begin{split}
T_n(u,v)=&2\sum_{r=0}^{n}\,\int_{0}^{u} ds\, s^{-2r} \int_{0}^{v}dw \, w \,\sum_{l=2r+1}^{\infty} \,\binom{l}{2r+1}\, a_l\, \left(\frac{s}{2}\right)^l\,\\
&\times \sum_{m=0}^{\infty}\sum_{j=0}^{\infty} s^m \, w^{2j+2n} \left(\frac{\mu}{2\hbar^2}\right)^{j+1} \alpha^{(n-r)}_{m+1,j+n-r}.
\end{split}
\end{equation} 

Now, note that the $n$th derivative of the potential $V(q)$ given by (\ref{nonlinear}) is
\begin{equation}\label{nthderi}
V^{(n)}(q)=n! \, \sum_{l=n}^{\infty} \, \binom{l}{n}\, a_l \, q^{l-n}.
\end{equation}
The above equation allows us to close the infinite series along $l$ in (\ref{tnuvapp6}), that is
\begin{equation} \label{vderi2r}
\sum_{l=2r+1}^{\infty} \,\binom{l}{2r+1}\, a_l\, \left(\frac{s}{2}\right)^l=\frac{1}{(2r+1)!}\left(\frac{s}{2}\right)^{2r+1}\, V^{(2r+1)}\left(\frac{s}{2}\right).
\end{equation}
Substituting (\ref{vderi2r}) into (\ref{tnuvapp6}), we arrive at
\begin{equation}\label{tnuvapp7}
\begin{split}
T_n(u,v)=&\left(\frac{\mu}{2\hbar^2}\right)\sum_{r=0}^{n}\,\frac{1}{(2r+1)!}\frac{1}{2^{2r}}\int_{0}^{u} ds\,  V^{(2r+1)}\left(\frac{s}{2}\right)\, \int_{0}^{v}dw \, w \,\\
&\times \sum_{m=0}^{\infty}\sum_{j=0}^{\infty} s^m \, w^{2j+2n} \left(\frac{\mu}{2\hbar^2}\right)^{j} \alpha^{(n-r)}_{m+1,j+n-r}.
\end{split}
\end{equation} 

Note that the coefficients $\alpha^{(n-r)}_{m+1,j+n-r}$ are nonvanishing only for $j \ge 1$. Performing a shift in indices from $m$ to $m-1$ and $j$ to $j+1$, and adding a factor $1=w^{2r-2r}$, we arrive at
\begin{equation}\label{tnuvapp8}
\begin{split}
T_n(u,v)=&\left(\frac{\mu}{2\hbar^2}\right)\sum_{r=0}^{n}\,\frac{1}{(2r+1)!}\frac{1}{2^{2r}}\int_{0}^{u} ds\,  V^{(2r+1)}\left(\frac{s}{2}\right)\, \int_{0}^{v}dw \, w^{2r+1} \,\\
&\times \sum_{m=0}^{\infty}\sum_{j=0}^{\infty} s^m \, w^{2j+2(n-r)+2} \left(\frac{\mu}{2\hbar^2}\right)^{j+1} \alpha^{(n-r)}_{m+1,j+(n-r)-1}.
	\end{split}
\end{equation} 
We compare the factor with double summation in the above equation with that of (\ref{tnuv}). The former is just the $n \to n-r$ case expressed in $(s,w)$ variables, that is,
\begin{equation}
T_{n-r}(s,w)=\sum_{m=0}^{\infty}\sum_{j=0}^{\infty} s^m \, w^{2j+2(n-r)+2} \left(\frac{\mu}{2\hbar^2}\right)^{j+1} \alpha^{(n-r)}_{m+1,j+(n-r)-1}.
\end{equation} 
Hence, the $n$th kernel factor correction $T_n(u,v)$ (\ref{tnuvapp8}) simplifies to
\begin{equation}\label{tnuvapp9}
\begin{split}
T_n(u,v)=&\left(\frac{\mu}{2\hbar^2}\right)\sum_{r=0}^{n}\,\frac{1}{(2r+1)!}\frac{1}{2^{2r}}\int_{0}^{u} ds\,  V^{(2r+1)}\left(\frac{s}{2}\right)\, \int_{0}^{v}dw \, w^{2r+1} \, T_{n-r}(s,w),       O
\end{split}
\end{equation} 
for $n\ge1$.

Again, we are able to rewrite the kernel factor correction for arbitrary $n$ as some integral of our interaction potential without explicitly solving for the expansion coefficients $\alpha^{(n)}_{m,j+n+1}$ in (\ref{tnuv}). 

Equation (\ref{tnuvapp8}) can be easily converted as a partial differential equation for $T_n(u,v)$ and is given by
\begin{equation}\label{tnuvpde}
\frac{\partial T_n(u,v)}{\partial v \partial u}=\left(\frac{\mu}{2\hbar^2}\right)\sum_{r=0}^{n}\frac{1}{(2r+1)!}\frac{1}{2^{2r}}  \, V^{(2r+1)} \left(\frac{u}{2}\right)\,v^{2r+1}\,T_{n-r}(u,v).
\end{equation}
subject to the boundary conditions $T_{n}(u,0)=T_{n}(0,v)=0$ in (\ref{bccorrect}). Equation (\ref{tnuvpde}) will be important later when we prove that the full solution $T(u,v)$ indeed satisfies the time kernel equation (\ref{tkeuv}). 

The existence of the solution of the time kernel equation guarantees the existence of $T_n(u,v)$. Similar to the partial differential equations for $T_0(u,v)$ and $T_1(u,v)$, it can be shown that the solution $T_n(u,v)$ is also unique with the boundary conditions specificed by (\ref{bccorrect}). The uniqueness will also follow from the use of the method of successive approximations on the solution $T_n(u,v)$ of the integral equation in (\ref{tnuvapp9}). Also, if we let $n=1$ in (\ref{tnuvapp9}) and (\ref{tnuvpde}), we recover correctly the integral and partial differential equations satisfied by the leading kernel correction $T_1(u,v)$.

We now go back to the integral equation given by (\ref{tnuvapp9}) and apply the same method we did in the previous subsections. We isolate the $r=0$ term in (\ref{tnuvapp9}) to get
\begin{equation}\label{app:tn1}
\begin{split}
T_n(u,v)&=\left(\frac{\mu}{2\hbar^2}\right) \int_{0}^{u} ds\, V^{(1)}\left(\frac{s}{2}\right)\int_{0}^{v}dw\,w\,T_n(s,w) \\
&+\left(\frac{\mu}{2\hbar^2}\right)\sum_{r=1}^{n} \frac{1}{(2r+1)!}\frac{1}{2^{2r}}\int_{0}^{u} ds \, V^{(2r+1)}\left(\frac{s}{2}\right)\int_{0}^{v}dw \, w^{2r+1}\,T_{n-r}(s,w). 
\end{split}
\end{equation}
We take the second term of (\ref{app:tn1}) as the zeroth order approximation of $T_n(u,v)$, that is,
\begin{equation}\label{app:tn3}
T_{n,0}(u,v)=\left(\frac{\mu}{2\hbar^2}\right)\sum_{r=1}^{n} \frac{1}{(2r+1)!}\frac{1}{2^{2r}}\int_{0}^{u} ds \, V^{(2r+1)}\left(\frac{s}{2}\right)\int_{0}^{v}dw \, w^{2r+1}\,T_{n-r}(s,w).
\end{equation}
The $m$th order approximation of the solution $T_n(u,v)$ is obtained from the following recurrence equation,
\begin{equation}\label{app:tn2}
T_{n,m}(u,v)=T_{n,0}(u,v)+\left(\frac{\mu}{2\hbar^2}\right)\int_{0}^{u} ds \, V'\left(\frac{s}{2}\right) \int_{0}^{v} dw \, w \, T_{n,m-1}(s,w),
\end{equation}
valid for $m \ge 1$. The kernel factor correction $T_n(u,v)$ can then be taken from the limit $T_n(u,v)=\lim\limits_{m\to\infty}T_{n,m-1}(u,v)$. 

The explicit forms of the first two iterations corresponding to $m=1,2$ are given by the following equations,
\begin{equation}\label{app:tn4}
\begin{split}
T_{n,1}(u,v)&=T_{n,0}(u,v)+\left(\frac{\mu}{2\hbar^2}\right)^2\sum_{r=1}^{n} \frac{1}{(2r+1)!}\frac{1}{2^{2r}}\int_{0}^{u} ds \, V^{(2r+1)}\left(\frac{s}{2}\right)\\
&\times \int_{0}^{v}dw \,w^{2r+1}\,(v^2-w^2) \, \left[V\left(\frac{u}{2}\right)-V\left(\frac{s}{2}\right)\right] \,T_{n-r}(s,w),
\end{split}
\end{equation}
\begin{equation}\label{app:tn5}
\begin{split}
T_{n,2}(u,v)&=T_{n,1}(u,v)+\left(\frac{\mu}{2\hbar^3}\right)^3\sum_{r=1}^{n} \frac{1}{(2r+1)!}\frac{1}{2^{2r}}\int_{0}^{u} ds \, V^{(2r+1)}\left(\frac{s}{2}\right)\\
&\times \int_{0}^{v}dw \,w^{2r+1}\,(v^2-w^2)^2 \, \left[V\left(\frac{u}{2}\right)-V\left(\frac{s}{2}\right)\right]^2 \,T_{n-r}(s,w).
\end{split}
\end{equation}

From the first few iterates of $T_{n,m}(u,v)$, we can infer the following general form for arbitrary $m$
\begin{equation}\label{app:tn6}
\begin{split}
T_{n,m}(u,v)&=\left(\frac{\mu}{2\hbar^2}\right)\sum_{r=1}^{n} \frac{1}{(2r+1)!}\frac{1}{2^{2r}}\int_{0}^{u} ds \, V^{(2r+1)}\left(\frac{s}{2}\right)\int_{0}^{v}dw \, w^{2r+1}\,T_{n-r}(s,w)\\
&\times\sum_{k=0}^{m}\frac{1}{(1)_kk!}\left(\frac{\mu}{2\hbar^2}\right)^k(v^2-w^2)^k\left[V\left(\frac{u}{2}\right)-V\left(\frac{s}{2}\right)\right]^k.
\end{split}
\end{equation}

Equation (\ref{app:tn6}) is formally proven by mathematical induction. Taking the limit $m\to \infty$, the sum along $k$ becomes a specific hypergeometric function. Hence, we finally find 
\begin{equation}\label{tnuvfin}
\begin{split}
T_n(u,v)=&\left(\frac{\mu}{2\hbar^2}\right)\sum_{r=1}^{n} \frac{1}{(2r+1)!}\frac{1}{2^{2r}}\int_{0}^{u} ds \, V^{(2r+1)}\left(\frac{s}{2}\right) \int_{0}^{v} dw \, w^{2r+1} \, T_{n-r}(s,w) \, \\
& \times {}_0F_1 \left(;1;\left(\frac{\mu}{2\hbar^2}\right)(v^2-w^2)\left[V \left(\frac{u}{2}\right)-V \left(\frac{s}{2}\right)\right]\right),
\end{split}
\end{equation}
valid for $n\ge 1$ which is our final expression for the $n$th kernel factor correction. The leading kernel $T_0(u,v)$ serves as the initial condition of the above equation. Equation (\ref{tnuvfin}) is in essence the main result of this study.

Letting $n=1$, we recover correctly the leading kernel correction $T_1(u,v)$ given by (\ref{t10uvfinb}). For the second kernel factor correction $T_2(u,v)$, letting $n=2$ leads to
\begin{equation}\label{t20uv}
\begin{split}
T_2(u,v)&=\frac{1}{4 \cdot 3!}\left(\frac{\mu}{2\hbar^2}\right)\int_{0}^{u} ds \, V^{(3)}\left(\frac{s}{2}\right) \int_{0}^{v} dw \, w^3 \, T_1(s,w) \, G(s,w)\\
&+\frac{1}{16 \cdot 5!}\left(\frac{\mu}{2\hbar^2}\right)\int_{0}^{u} ds \, V^{(5)}\left(\frac{s}{2}\right) \int_{0}^{v} dw \, w^5 \, T_0(s,w) \, G(s,w)
\end{split}
\end{equation}
where
\begin{equation}
G(s,w)={}_0F_1 \left(;1;\left(\frac{\mu}{2\hbar^2}\right)(v^2-w^2)\left[V \left(\frac{u}{2}\right)-V \left(\frac{s}{2}\right)\right]\right).
\end{equation}

Both the kernel factors $T_0(u,v)$ and $T_1(u,v)$ are already known at this point so that $T_2(u,v)$ can also be solved analytically or numerically for a given potential $V(q)$. Likewise, the $n=3$ case leads to the third kernel correction $T_3(u,v)$ given by 
\begin{equation}\label{t30uv}
\begin{split}
T_3(u,v)&=\frac{1}{4 \cdot 3!}\left(\frac{\mu}{2\hbar^2}\right)\int_{0}^{u} ds \, V^{(3)}\left(\frac{s}{2}\right) \int_{0}^{v} dw \, w^3 \, T_2(s,w) \, G(s,w)\\
&+\frac{1}{16 \cdot 5!}\left(\frac{\mu}{2\hbar^2}\right)\int_{0}^{u} ds \, V^{(5)}\left(\frac{s}{2}\right) \int_{0}^{v} dw \, w^5 \, T_1(s,w) \, G(s,w)\\
&+\frac{1}{64 \cdot 7!}\left(\frac{\mu}{2\hbar^2}\right)\int_{0}^{u} ds \, V^{(7)}\left(\frac{s}{2}\right) \int_{0}^{v} dw \, w^7 \, T_0(s,w) \, G(s,w).
\end{split}
\end{equation}

All the higher order corrections can similarly be determined from (\ref{tnuvfin}) so that in principle, we now have a complete supraquantized TOA-operator in accordance with (\ref{toadef})-(\ref{timekernel}). 
\subsection{Properties of the kernel factor corrections}
We now determine some important properties of the kernel factor corrections $T_n(u,v)$ given by (\ref{tnuvfin}).

\vspace{1mm}

\textit{1. Vanishing of $T_n(u,v)$ for $u$ or $v \to 0$}. 

\vspace{1mm}
We can see immediately that $T_n(u,v)$ satisfies the boundary conditions $T_n(u,0)=0$ and $T_n(0,v)=0$ so that the full time kernel factor $T(u,v)$ satisfies the original boundary conditions given by (\ref{boundary}). The vanishing of the corrections for $u$ or $v \to 0$ guarantees that the supraquantized TOA-operator leads to the correct classical arrival time in the classical limit. 

\vspace{1mm}

\textit{2. Symmetry of $T_n(u,v)$ along $v$}. 

\vspace{1mm}

Another important property of the corrections $T_n(u,v)$ is its symmetry along $v$. Changing variables from $v$ to $-v$ in (\ref{tnuvfin}) and noting that the initial condition $T_0(u,v)$ is symmetric along $v$, we arrive at the relation
$T_n(u,v)=T_n(u,-v)$. This guarantees that the expectation value of the corresponding operator, the quantum correction  $\mathrm{\hat{T}}_n$, is always real valued.  

\vspace{1mm}

\textit{3. Vanishing and non-vanishing of $T_n(u,v)$ for linear and nonlinear systems, respectively.} 

\vspace{1mm}

Probably the most significant property of the time kernel factor corrections $T_n(u,v)$ is its vanishing for linear systems and non-vanishing for nonlinear systems. For linear systems of the form $V(q)=a+bq+cq^2$, the factor $V^{(2r+1)}\left(s/2\right)$ in (\ref{tnuvfin}) is always zero for $r\ge 1$. Hence, there are no quantum corrections to the Weyl quantization of the classical time of arrival for this case. This is the exact reason why the Weyl-quantized TOA-operator is sufficient for specific quantum arrival time problems involving a free-particle, a particle in a gravitational field and a particle in the presence of harmonic oscillator, among others. On the other hand, the factor $V^{(2r+1)}\left(s/2\right)$ in (\ref{tnuvfin}) is always non-zero for the case of nonlinear systems. This explains why there always exist quantum corrections to the Weyl-quantized TOA-operator for nonlinear systems. These observations clearly explain why the Weyl-quantized TOA-operator satisfy the conjugacy requirement with the system Hamiltonian for linear systems but not for nonlinear systems. Equivalently, the existence of obstruction to quantization in the quantum time of arrival problem is justified by the appearance of these quantum corrections to the Weyl-quantization of the classical arrival time. 

\vspace{1mm}

\textit{4. Dependence of the corresponding kernel $\langle q|\hat{T}_n|q'\rangle$ on $\hbar^{2n}$.} 

\vspace{1mm}

We first calculate the general form of the Weyl-Wigner transform of the $n$th kernel factor correction. Using the definition of $T_n(u,v)$ in (\ref{tnuvapp}), the corresponding time kernel is given by
\begin{equation}
\begin{split}
\left\langle q\left|T_n\right|q'\right\rangle=\frac{\mu}{i\hbar}\,\mathrm{sgn}(q-q')\,\sum_{m=0}^{\infty}\sum_{j=0}^{\infty}\,(q+q')^m \,(q-q')^{2j+2n+2}\left(\frac{\mu}{2 \hbar^2}\right)^{j+1}\alpha^{(n)}_{m,j+n+1},
\end{split}
\end{equation}
where the coefficients $\alpha^{(n)}_{m,j+n+1}$ satisfy recurrence relation given by (\ref{alphasmj}). Its Weyl-Wigner transform is given by
\begin{equation}\label{weylwignertn}
\begin{split}
\mathcal{T}_{n}(q,p)&=\int_{-\infty}^{\infty} \, d\nu \,\left\langle q+\frac{\nu}{2}\left|T_n\right|q-\frac{\nu}{2}\right\rangle \mathrm{e}^{-i \nu p/\hbar}\\
&=2\,\frac{\mu}{p^{3+2n}}\,\hbar^{2n} \,  \sum_{m=0}^{\infty}\sum_{j=0}^{\infty}\,(-1)^{j+n}\,(2q)^m \,(2j+2n+2)! \left(\frac{\mu}{2p^2}\right)^{j+1}\alpha_{m,j+n+1}^{(n)}.
\end{split}
\end{equation}
Equation (\ref{weylwignertn}) can be converted as an integral equation by performing similar series rearrangement as in the previous subsections. The result is given by
\begin{equation}\label{wwtn}
\mathcal{T}_{n}(q,p)=\frac{\mu}{p}\,\sum_{r=0}^{n}\,\hbar^{2r}\,\frac{(-1)^r}{2^{2r}\,(2r+1)!} \,\int_{0}^{q}dq'\, V^{(2r+1)}(q')\, \frac{\partial^{2r+1} \, \mathcal{T}_{n-r}(q',p) }{\partial p^{2r+1}}.
\end{equation}
Solving (\ref{wwtn}) by the same method of successive approximation as we did before, we arrive at
\begin{equation}\label{wwtnfin}
\mathcal{T}_{n}(q,p)=\mu\,\hbar^{2n}\,\sum_{r=1}^{n}\,\frac{(-1)^r}{2^{2r}\,(2r+1)!} \,\int_{0}^{q}dq'\,\,\mathrm{exp}\left[\left(V(q)-V(q')\right)\,\frac{\mu}{p}\frac{\partial}{\partial\,p}\right]\frac{1}{p} \,V^{(2r+1)}(q')\, \frac{\partial^{2r+1} \, \mathcal{T}_{n-r}(q',p) }{\partial p^{2r+1}}.
\end{equation}
for $n\ge1$. The Weyl-Wigner transform $\mathcal{T}_{n}(q,p)$ is also vanishing (nonvanishing) for linear (nonlinear) systems due to the vanishing (nonvanishing) of the factor $V^{(2r+1)}(q')$ in (\ref{wwtnfin}). 

We now clearly see the explicit dependence of $\mathcal{T}_{n}(q,p)$ on $\hbar^{2n}$. This vanishes in the classical limit $\hbar \to 0$ so that the supraquantized TOA-operator leads to the classical arrival time and hence satisfies the quantum-classical correspondence principle. 

In essense, the $\mathcal{T}_{n}(q,p)$'s can also be regarded as the quantum corrections to the classical arrival time in phase space so that quantizing them using Weyl prescription leads to the time kernel corrections $\langle q|\hat{T}_n|q'\rangle$.
\vspace{1mm}

\textit{5. Decreasing contribution of $T_n(q,q')$ with increasing $n$}. 

\vspace{1mm}

Numerical evaluations of $T_n(u,v)$ for some specific values of $u$, $v$ and $V(q)$ imply a decreasing contribution of $T_n(q,q')$ with increasing $n$. This can already be expected since its Weyl-Wigner transform (\ref{wwtnfin}) clearly illustrates the explicit dependence on $\hbar^{2n}$ which has a decreasing classical contribution with $n$. Hence, it does make sense to approximate the full time kernel factor $T(q,q')$ as a partial sum $T(q,q')\approx T_0(q,q')+T_1(q,q')+...+T_m(q,q')$, with the approximation getting more accurate as more terms are added. 

One may argue that the quantum corrections are too small and can be neglected right away in any calculations so that approximating the supraquantized TOA-operator as just the Weyl-quantized TOA-operator suffices. However, the choice to neglect the quantum corrections should depend strictly on the specific quantum time of arrival problem being considered. For example, we consider the case of quantum tunneling. In Ref. \cite{Galapon2012}, 
the leading term of the supraquantized TOA-operator (\ref{supraexpand}) for the barrier case was only considered. Its expectation value is found to be vanishing for the case of quantum tunneling. However, the zero contribution from the leading term does not immediately mean that the expectation values of the quantum corrections identically vanish. In the same vein, the experimental attosecond time measurements implying zero quantum tunneling time are strictly valid up to the attosecond regime only\cite{Eckle2008,Pfeiffer2011,Sainadh2019}. We do not know if the same result extends to the zeptosecond regime. This is exactly one of our motivations why the quantum corrections need to be explicitly determined and the supraquantized TOA-operator needs to be completed. 

\subsection{The complete time kernel factor and the supraquantized TOA-operator}
To completely and finally validate our results, we now show that the full solution 
\begin{equation}\label{expandtuv}
T(u,v)=T_0(u,v)+\sum_{n=1}^{\infty}T_n(u,v),
\end{equation}
where $T_0(u,v)$ and $T_n(u,v)$ are defined by (\ref{t0uvfin}) and $(\ref{tnuvfin})$, respectively,  indeed satisfies the time kernel equation in (\ref{tkeuv}), that is, \begin{equation}\label{tkesat1}
\frac{2\hbar^2}{\mu}\frac{\partial^2 \,T(u,v)}{\partial u \,\partial v}=\left(V\left(\frac{u+v}{2}\right)-V\left(\frac{u-v}{2}\right)\right)T(u,v).
\end{equation}

The form of $T(u,v)$ as indicated by (\ref{expandtuv}) implies that the left-hand side of the time kernel equation (\ref{tkesat1}) involves the partial derivatives $\partial^2 \,T_0(u,v)/\partial u\,\partial v$ and $\partial^2 \,T_n(u,v)/\partial u\,\partial v$. But these two derivatives are exactly the partial differential equations uniquely satisfied by the kernel factors $T_0(u,v)$ and $T_n(u,v)$ as shown in (\ref{t0uvpde}) and (\ref{tnuvpde}), respectively. The left-hand side of (\ref{tkesat1}) then evaluates to
\begin{equation}\label{tkesat2}
\begin{split}
\frac{2\hbar^2}{\mu}\frac{\partial^2 \,T(u,v)}{\partial u \,\partial v}&=\sum_{n=0}^{\infty} \sum_{r=0}^{n}\frac{1}{(2r+1)!}\frac{1}{2^{2r}}  \, V^{(2r+1)} \left(\frac{u}{2}\right)\,v^{2r+1}\,T_{n-r}(u,v).
\end{split}
\end{equation}
Using the summation identity in (\ref{identitysum}), the two summations can be decoupled leading to 
\begin{equation}\label{tkesat3}
\frac{2\hbar^2}{\mu}\frac{\partial^2 \,T(u,v)}{\partial u \,\partial v}=\sum_{r=0}^{\infty} \frac{1}{(2r+1)!}\frac{1}{2^{2r}}  \, V^{(2r+1)} \left(\frac{u}{2}\right)\,v^{2r+1}\,T(u,v).
\end{equation}

Now, the kernel factor $T(u,v)$ satisfies the time kernel equation if the right hand sides of (\ref{tkesat1}) and (\ref{tkesat3}) are equal, which we will explicitly show. To proceed, we need to evaluate the right-hand side of (\ref{tkesat1}). For entire analytic potentials of the form $V(q)=\sum_{l=1}^{\infty}a_l\, q^l$ and noting the binomial expansion
\begin{equation}
(x+y)^l=\sum_{m=0}^{l} \, \binom{l}{m}\,x^{l-m}\,y^m,
\end{equation} 
we find the following equality
\begin{equation}\label{tkesat4}
\begin{split}
\left(V\left(\frac{u+v}{2}\right)-V\left(\frac{u-v}{2}\right)\right)T(u,v)=\sum_{l=1}^{\infty}\,\frac{a_l}{2^{l-1}}\,\sum_{m=0}^{[l]} \binom{l}{2m+1}u^{l-2m-1}v^{2m+1} \,T(u,v),
\end{split}
\end{equation}
where the index $[l]$ is defined as $[l]=(l-1)/2$ for odd $l$ and $[l]=l/2-1$ for even $l$. 

We separate the even and odd parts of the sum along $l$, and then perform a shift in index from $l$ to $l+1$. Doing so leads us to
\begin{equation}\label{tkesat5}
\begin{split}
\left(V\left(\frac{u+v}{2}\right)-V\left(\frac{u-v}{2}\right)\right)T(u,v)&=\sum_{l=0}^{\infty}\sum_{m=0}^{l}\,\frac{a_{2l+1}}{2^{2l}}\, \binom{2l+1}{2m+1}u^{2l-2m}v^{2m+1} \,T(u,v)\\
&+\sum_{l=0}^{\infty}\sum_{m=0}^{l}\,\frac{a_{2l+2}}{2^{2l+1}}\, \binom{2l+2}{2m+1}u^{2l-2m+1}v^{2m+1} \,T(u,v).
\end{split}
\end{equation}
Using again the summation identity in (\ref{identitysum}) followed by a shift of index from $l$ to $l-(2m+1)$, (\ref{tkesat5}) simplifies to
\begin{equation}\label{tkesat6}
\begin{split}
\left(V\left(\frac{u+v}{2}\right)-V\left(\frac{u-v}{2}\right)\right)T(u,v)&=\sum_{m=0}^{\infty}\frac{v^{2m+1}}{2^{2m}} \sum_{l=2m+1}^{\infty}a_l\, \binom{l}{2m+1}\left(\frac{u}{2}\right)^{l-(2m+1)}\,T(u,v).\\
\end{split}
\end{equation}

Now, the sum along $l$ is related to the $(2m+1)$th derivative of the potential $V(q)$ evaluated at $q=u/2$, that is,
\begin{equation}
V^{(2m+1)}\left(\frac{u}{2}\right)=(2m+1)!\sum_{l=2m+1}^{\infty}a_l\, \binom{l}{2m+1}\left(\frac{u}{2}\right)^{l-(2m+1)}.
\end{equation}
Hence, we finally arrive at the following result
\begin{equation}\label{tkesat7}
\left(V\left(\frac{u+v}{2}\right)-V\left(\frac{u-v}{2}\right)\right)T(u,v)=\sum_{m=0}^{\infty} \frac{1}{(2m+1)!}\frac{1}{2^{2m}}  \, V^{(2m+1)} \left(\frac{u}{2}\right)\,v^{2m+1}\,T(u,v).
\end{equation}

We immediately see that the right-hand side of (\ref{tkesat7}) is equal to the right-hand side of (\ref{tkesat3}). Equating both equations, we immediately arrive at the time kernel equation given by (\ref{tkesat1}). Hence, the full solution $T(u,v)$, with the leading term $T_0(u,v)$ and the succeeding terms $T_n(u,v)$ for $n\ge1$, satisfies the time kernel equation. This validates our results for the kernel factor corrections $T_n(u,v)$ and consequently the quantum corrections $\hat{\mathrm{T}}_n$. 


\section{The TOA problem for the quartic anharmonic oscillator}\label{sec:examples}

We now apply our results to a specific quantum system, in particular, the case of quartic anhamonic oscillator of the form $V(q)=\lambda q^4$ for some constant $\lambda$. This potential clearly yields a nonlinear equation of motion.

\subsection{The Weyl-quantized TOA-operator}
With the given potential, the leading kernel factor $T_0(u,v)$ in accordance to (\ref{t10uvfinb}) is given by
\begin{equation}
T_0(u,v)=\frac{1}{4}\int_{0}^{u} ds \,\, {}_0F_1\left(;1;\, \eta \, v^2 \,  (u^4-s^4)\right),
\end{equation}
where $\eta=\mu \lambda/32\hbar^2.$ Expanding the integrand and exchanging the order of integration and summation, we arrive at the following equation
\begin{equation}\label{t0anharm}
T_0(u,v)=\frac{1}{4}\sum_{k=0}^{\infty}\frac{(\eta \, v^2)^k}{(1)_k \, k!} \int_{0}^{u}ds \,  (u^4-s^4)^k.
\end{equation}
The interchange is valid due to the absolute convergence of the hypergeometric function ${}_0F_{1}(;1;z)$. 

The integral along $s$ in (\ref{t0anharm}) is easily evaluated and is given by
\begin{equation}
\int_{0}^{u}ds \,  (u^4-s^4)^k= u^{4k+1} \, \frac{\Gamma\left(\frac{5}{4}\right)\Gamma\left(k+1\right)}{\Gamma\left(k+\frac{5}{4}\right)},
\end{equation}
so that the leading time kernel factor assumes the form
\begin{equation}\label{t0anharm2}
T_0(u,v)=\frac{u}{4}\sum_{k=0}^{\infty}\frac{(\eta\, u^4 \, v^2 )^k}{(1)_k} \, \frac{\Gamma\left(\frac{5}{4}\right)}{\Gamma\left(k+\frac{5}{4}\right)}.
\end{equation}
Equation (\ref{t0anharm2}) can be summed in closed-form in terms of a specific hypergeometric function leading to
\begin{equation}\label{t0anharm4}
T_0(u,v)=\frac{u}{4} {}_0F_1 \left(;\frac{5}{4};\left(\frac{\mu \lambda}{32\hbar^2}\right)u^4 v^2\right).
\end{equation}

The above result is also the Weyl-quantized time kernel factor for the anharmonic potential. One could show that $T_0(u,v)$ satisfies the partial differential equation for the leading kernel factor (\ref{t0uvpde}), that is,
\begin{equation}\label{leadingpdeanharm}
\frac{\partial^2 \, T_0(u,v)}{\partial v  \, \partial u}= 8 \, \eta \, u\,  v^3 \, T_0(u,v),
\end{equation}
and the boundary conditions $T_0(u,0)=u/4$ and $T_0(0,v)=0$. Hence, (\ref{t0anharm4}) is validated.

The leading kernel of the supraquantized TOA-operator (\ref{toadef}) for the anharmonic oscillator is then given by
\begin{equation}\label{leadingkernqua}
\left\langle q\left|T_0\right|q'\right\rangle=\frac{\mu}{i\hbar} \,\mathrm{sgn}(q-q')\,\frac{q+q'}{4}\, {}_0F_1 \left(;\frac{5}{4};\left(\frac{\mu \lambda}{32\hbar^2}\right)(q+q')^4\, (q-q')^2\right),
\end{equation}
which also coincides with the time kernel of the Weyl-quantized TOA-operator. Taking the Weyl-Wigner transform of (\ref{leadingkernqua}) in accordance with (\ref{oinverse}), we have
\begin{equation}
\begin{split}
\tau_0(q,p)&=\int_{-\infty}^{\infty}d\nu \,\left\langle q+\frac{\nu}{2}\left|T_0\right|q-\frac{\nu}{2}\right\rangle \mathrm{e}^{-i \nu p/\hbar}\\
&=\frac{\mu q}{2i\hbar}\int_{-\infty}^{\infty}d\nu\,\, {}_0F_1 \left(;\frac{5}{4};\left(\frac{\mu \lambda q^4}{2\hbar^2}\right) \nu^2\right)\,\mathrm{sgn}(\nu)\,\mathrm{e}^{-ip\nu\hbar}
\end{split}
\end{equation}

The above integral can be evaluated by expanding the hypergeometric function as an infinite series, exchanging the order of summation and integration, and then integrating term by term using the integral identity in (\ref{seventh}). The result is given by
\begin{equation}\label{tau0quartic}
\tau_0(q,p)=-\frac{\mu q}{p}\,\,{}_2F_1\left(\frac{1}{2},1;\frac{5}{4};-\frac{2\mu\lambda q^4}{p^2}\right),
\end{equation}
which is clearly independent of $\hbar$. 

Equation (\ref{tau0quartic}) is also the classical time of arrival of a particle at the origin in the presence of a quartic anharmonic oscillator potential. This can be checked by substituting our potential into the general classical time of arrival expression in (\ref{classical}). Note that $\tau_0(q,p)$ is strictly positive since the initial position $q$ is located to the left of the origin. It is interesting to note that the classical time of arrival appears as a free-partical arrival time $\tau_F(q,p)=-\mu q/p$ deformed by some function dependent on the potential $V(q)=\lambda q^4$.

\subsection{The first three leading kernel factor corrections}

Let us now calculate the first three quantum corrections  (\ref{tnuvfin}) to the Weyl quantization of the classical arrival time for the case of a quartic anharmonic oscillator potential. This is done by explicitly solving for the first three kernel factor corrections $T_1(u,v), T_2(u,v)$, and $T_3(u,v)$. 

In (\ref{tnuvfin}), the factor $V^{(2r+1)}\left(s/2\right)$ is vanishing for $r \ge 2$ so that only the $r=1$ term contributes in  the sum. The $n$th quantum correction then assumes the form
\begin{equation}\label{nthcorrectionquart}
\begin{split}
T_n(u,v)=&\frac{\mu}{48\hbar^2}\int_{0}^{u} ds \, V^{(3)}\left(\frac{s}{2}\right) \int_{0}^{v} dw \, w^{3} \, T_{n-1}(s,w) \,  {}_0F_1 \left(;1;\left(\frac{\mu}{2\hbar^2}\right)(v^2-w^2)\left[V \left(\frac{u}{2}\right)-V \left(\frac{s}{2}\right)\right]\right),
\end{split}
\end{equation} 
for all $n\ge1$. Substituting our potential $V(q)$ and the leading kernel factor $T_0(u,v)$ into (\ref{nthcorrectionquart}), the leading kernel factor correction is given by 
\begin{equation}\label{t10uvfinanharm3}
\begin{split}
T_1(u,v)=2 \,\eta\int_{0}^{u} ds \, \int_{0}^{v} dw \,  w^3 \, s^2 \, {}_0F_1 \left(;\frac{5}{4};\eta \, s^4 w^2\right)\,{}_0F_1 \left(;1;\eta \, (v^2-w^2)(u^4-s^4)\right),
\end{split}
\end{equation}
where $\eta=\mu \lambda/32\hbar^2.$ 

To evaluate the double integral, we expand the two hypergeometric functions in the integrand, and then interchange the order of summations and integrations. As a result, we find
\begin{equation}\label{t10uvfinanharm4}
\begin{split}
T_1(u,v)= 2 \, \eta \sum_{k=0}^{\infty}\sum_{l=0}^{\infty} \frac{\eta^{k+l}}{(5/4)_k (1)_l} \frac{1}{k! \, l!} \int_{0}^{u} ds \, s^{4k+2}\,(u^4-s^4)^l \int_{0}^{v} dw \,  \,  w^{2k+3}\,(v^2-w^2)^l.
\end{split}
\end{equation}
The integrals along $s$ and $w$ can be straightforwardly evaluated leading to
\begin{equation}\label{integralanharm}
\int_{0}^{u} ds \, s^{4k+2}\,(u^4-s^4)^l= u^{4k+4l+3} \, \frac{\Gamma(k+3/4)\,\Gamma(l+1)}{4 \,  \Gamma(k+l+7/4)},
\end{equation}
\begin{equation}\label{integralanharm2}
\int_{0}^{v} dw \,  \,  w^{2k+3}\,(v^2-w^2)^l=v^{2k+2l+4} \,  \frac{\Gamma(k+2)\,\Gamma(l+1)}{2 \, \Gamma(k+l+3)}.
\end{equation}
Substituting (\ref{integralanharm}) and (\ref{integralanharm2}) into (\ref{t10uvfinanharm4}), our leading kernel factor correction $T_1(u,v)$ becomes
\begin{equation}\label{t10uvfinanharm5}
T_1(u,v)=\frac{\Gamma(5/4)}{4} \, \eta \, u^3 \, v^4\, \sum_{k=0}^{\infty} \sum_{l=0}^{\infty} \left(\eta \, v^2 \, u^4 \right)^{k+l} \, \frac{(k+1) \, \Gamma(k+3/4)}{\Gamma(k+5/4) \, \Gamma(k+l+7/4) \, \Gamma(k+l+3)}.
\end{equation}

We rearrange the two summations using the following identity \cite{Srivastava1984a}, 
\begin{equation}\label{identity}
\sum_{n=0}^{\infty} \sum_{m=0}^{\infty}A(m,n)=\sum_{n=0}^{\infty} \sum_{m=0}^{n} A(m,n-m),
\end{equation}
so that (\ref{t10uvfinanharm5}) becomes 
\begin{equation}\label{t10uvfinanharm6}
T_1(u,v)=\frac{\Gamma(5/4)}{4} \, \eta \, u^3\, v^4 \,  \sum_{l=0}^{\infty}  \frac{\left(\eta \, v^2 \, w^4 \right)^l}{\Gamma(l+7/4)\,\Gamma(l+3)}\sum_{k=0}^{l}  \, \frac{(k+1) \, \Gamma(k+3/4)}{\Gamma(k+5/4)}.
\end{equation}

The sum along $k$ results to a specific gamma function which enables us to sum the infinite series along $l$ into a specific hypergeometric function. The leading kernel factor correction is simply given by
\begin{equation}\label{t10uvfinanharm8}
T_1(u,v)=\frac{\eta \, u^3\,v^4}{24} \,  \left[5\,{}_2F_3\left(1,\frac{7}{2};\frac{5}{4},\frac{5}{2},3;\, \eta \, u^4\, v^2\right) - {}_1F_2 \left(1;\frac{7}{4},3;\, \eta\, u^4\,v^2 \right)\right].
\end{equation}

Using (\ref{nthcorrectionquart}) and following similar steps as before, the next two corrections are found to be
\begin{equation}
\begin{split}
T_2(u,v)=&\frac{\eta^2 u^5 v^8}{540}\left[{}_3F_4\left(2,2,2;1,1,\frac{9}{4},5;\eta u^4v^2\right) + \frac{43}{2}\, {}_2F_3\left(2,\frac{113}{27};\frac{9}{4},\frac{86}{27},5;\eta u^4v^2\right)\right.\\
&-\left.\frac{75}{8}\, {}_2F_3\left(1,\frac{17}{2};\frac{7}{4},5,\frac{15}{2};\eta u^4v^2\right)+\frac{39}{8}\, {}_1F_2\left(1;\frac{9}{4},5;\eta u^4v^2\right)\right],
\end{split}
\end{equation}
\begin{equation}
\begin{split}
T_3(u,v)=&\frac{\eta^3 u^7 v^{12}}{56700}\left[\frac{53}{9}\,{}_4F_5\left(2,2,2,\frac{60}{7};1,1,\frac{9}{4},7,\frac{53}{7};\eta u^4 v^2\right)-\frac{7}{15}\,{}_1F_2\left(1;\frac{9}{4},7;\eta u^4v^2\right)\right.\\
&+\frac{27}{4}\, {}_3F_4\left(2,2,2;1,1,\frac{9}{4},7;\eta u^4 v^2\right)-\frac{5}{6}\, {}_3F_4\left(2,2,2;1,1,\frac{11}{4},7;\eta u^4 v^2\right)\\
&+\frac{4921}{72}\, {}_2F_3\left(2,2;1,\frac{9}{4},7;\eta u^4v^2\right)+\frac{8633}{80}\, {}_2F_3\left(1,\frac{21277}{12644};\frac{8633}{12644},\frac{9}{4},7;\eta u^4v^2\right)\\
&-\frac{1115}{12}\, {}_2F_3\left(2,\frac{515}{69};\frac{11}{4},\frac{446}{69},7;\eta u^4v^2\right)+\frac{2275}{1600}\, {}_2F_3\left(1,\frac{27}{2};\frac{9}{4},7,\frac{25}{2};\eta u^4v^2\right)\\
&\left.-\frac{1375}{24}\,{}_1F_2\left(1;\frac{11}{4},7;\eta u^4v^2\right)+\frac{19}{45}\,{}_1F_2\left(2;\frac{9}{4},7;\eta u^4v^2\right)\right].
\end{split}
\end{equation}

The above results are validated by showing that each correction satisfies the partial differential equation for $T_n(u,v)$ (\ref{tnuvpde}), that is,
\begin{equation}\label{1stpdeanharm}
\frac{\partial^2 \, T_n(u,v)}{\partial v  \, \partial u}=8 \, \eta \, u^3 \, v \,  T_n(u,v)+ 8 \, \eta \, u\, v^3 \, T_{n-1}(u,v).
\end{equation}
and the boundary conditions $T_n(u,0)=0$ and $T_n(0,v)=0$. 

Clearly, the kernel factor corrections $T_1(u,v)$, $T_2(u,v)$, and $T_3(u,v)$ are non-vanishing and provide corrections to the Weyl-quantized time kernel factor $T_0(u,v)$. In addition, numerical evaluations of $T_0(u,v)$, $T_1(u,v)$, $T_2(u,v)$, and $T_3(u,v)$ implies that $T_3(u,v)<T_2(u,v)<T_1(u,v)<T_0(u,v)$. This supports our earlier assertion that the kernel factor corrections $T_n(u,v)'s$ have decreasing contribution with $n$. For this specific case, it is not unreasonable to expect the full time kernel factor $T(q,q')$ can be approximated by the partial sum $\sum_{n=0}^{3}T_n(q,q')$ so that the supraquantized-TOA operator for the quartic anharmonic oscillator is approximated up to the third order correction. The approximation gets better as we add more terms in the partial sum. 

An advantage of our results here is that we are able to solve the time kernel factor $T_n(u,v)$ up to the third order approximation analytically, unlike before where we have no choice but just to approximate the supraquantized TOA-operator using just the leading term. In fact, higher order corrections can still be determined, albeit tediously. The need to continue approximating our time kernel factor depends on how accurate we want our supraquantized TOA-operator to be when compared to the corresponding experimental TOA observable. 

For completeness, the Weyl-Wigner transforms of the kernel factor corrections are computed and are given by
\begin{equation}
\mathcal{T}_1(q,p)=-\mu^2 \lambda\,\frac{q^3}{p^2}\,\hbar^2\left[\frac{5}{2}\,\,{}_2F_1\left(1,\frac{7}{2};\frac{5}{4};-\frac{2\mu\lambda q^4}{p^2}\right)-\frac{1}{2}\,\,{}_2F_1\left(1,\frac{5}{2};\frac{7}{4};-\frac{2\mu\lambda q^4}{p^2}\right)\right],
\end{equation}
\begin{equation}
\begin{split}
\mathcal{T}_2(q,p)=-\mu^3 \lambda^2\,\frac{q^5}{p^9}\,\hbar^4&\,\,\left[\frac{14}{3}\,\, {}_4F_3\left(2,2,2,\frac{9}{2}\,;1,1,\frac{9}{4}; -\frac{2\mu \lambda\,q^4}{p^2}\right)+\frac{301}{3}\, {}_3F_2\left(2,\frac{113}{27},\frac{9}{2}\,;\frac{9}{4},\frac{86}{37}; -\frac{2\mu \lambda\,q^4}{p^2}\right)\right.\\
&-\left.\frac{175}{4}\,\, {}_3F_2\left(1,\frac{9}{2},\frac{17}{2}\,;\frac{7}{4},\frac{15}{2}\,;\, -\frac{2\mu \lambda\,q^4}{p^2}\right)+\frac{91}{4}\,\, {}_2F_1\left(1,\frac{9}{2}\,\,;\,\frac{9}{4}\,;-\frac{2\mu \lambda\,q^4}{p^2}\right)\right],
\end{split}
\end{equation}
\begin{equation}
\begin{split}
\mathcal{T}_3(q,p)=&-\mu^4 \lambda^3\frac{q^7}{p^{13}}\hbar^6\left[\frac{3498}{49}{}_5F_4\left(2,2,2,\frac{60}{7},\frac{13}{2};1,1,\frac{9}{4},\frac{53}{7}-\frac{2\mu \lambda q^4}{p^2}\right)-\frac{154}{5}{}_2F_1\left(1,\frac{13}{2};\frac{9}{4};-\frac{2\mu \lambda q^4}{p^2}\right)\right.\\
&+\frac{891}{2}\, {}_4F_3\left(2,2,2,\frac{13}{2};1,1,\frac{9}{4};-\frac{2\mu \lambda q^4}{p^2}\right)-55\, {}_4F_3\left(2,2,2,\frac{13}{2};1,1,\frac{11}{4};-\frac{2\mu \lambda q^4}{p^2}\right)\\
&+\frac{54131}{12}\, {}_3F_2\left(2,2,\frac{13}{2};1,\frac{9}{4};-\frac{2\mu \lambda q^4}{p^2}\right)+\frac{284889}{40}\, {}_3F_2\left(1,\frac{21277}{12644},\frac{13}{2};\frac{8633}{12644},\frac{9}{4};-\frac{2\mu \lambda q^4}{p^2}\right)\\
&-\frac{12265}{2}\, {}_3F_2\left(2,\frac{515}{69},\frac{13}{2};\frac{11}{4},\frac{446}{69};-\frac{2\mu \lambda q^4}{p^2}\right)+\frac{3003}{32}\, {}_3F_2\left(1,\frac{13}{2},\frac{27}{2};\frac{9}{4},\frac{25}{2};-\frac{2\mu \lambda q^4}{p^2}\right)\\
&\left.-\frac{15125}{24}\,{}_2F_1\left(1,\frac{13}{2};\frac{11}{4};-\frac{2\mu \lambda q^4}{p^2}\right)+\frac{418}{15}\,{}_2F_1\left(2,\frac{13}{2};\frac{9}{4};-\frac{2\mu \lambda q^4}{p^2}\right)\right].
\end{split}
\end{equation}
We see the explicit $\hbar^2$, $\hbar^4$, and $\hbar^6$ dependence of $\mathcal{T}_1(q,p)$, $\mathcal{T}_2(q,p)$, and $\mathcal{T}_3(q,p)$ respectively. In the classical limit $\hbar \to 0$, their contributions identically vanish so that the supraquantized TOA-operator leads to the correct classical time of arrival given by (\ref{tau0quartic}). Of course, the correction terms are required so that the complementary relation with the Hamiltonian strictly holds. 

Having constructed the supraquantized TOA-operator up to the third correction, the corresponding physical contents of the operator for the case of anharmonic oscillator potential may already be investigated in the standard way. However, the mathematical construction of the quantum corrections and their implementation to a specific quantum system already suffice for our current purposes. Elsewhere, the physical contents and implications of the quantum corrections are investigated by considering more realistic potentials that can be compared to experimental time of arrival measurements.

\section{Conclusion}\label{sec:conclusion}

In this paper, we have determined explicitly the quantum corrections to the Weyl-quantized time of arrival operator and expressed them as some integrals of the interaction potential. These corrections arise by imposing strict conjugacy of our supraquantized TOA-operator with the system Hamiltonian. They always vanish for linear systems but generally nonvanishing for the case of nonlinear systems. We then considered the case of quartic anharmonic oscillator potentials where we have computed the corresponding Weyl-quantized TOA-operator and the three leading kernel factor corrections. We showed that the Weyl-Wigner transform of the quantum corrections identically vanish in the classical limit $\hbar \to 0$. At this moment, we now have a complete supraquantized time of arrival operator for arbitrary entire analytic potentials which satisfies all important properties of a time of arrival observable such as the quantum-classical correspondence principle, time reversal symmetry, hermiticity, and conjugacy with the hamiltonian. Expectation values, eigenvalues, eigenfunctions, and probability distributions are constructed from the supraquantized TOA-operator in the standard way. Elsewhere, we will use the results obtained in this paper to investigate the quantum tunneling time of an elementary particle through piecewise rectangular and smooth potential barriers and to explore the exact role of the time-energy canonical commutation relation to the dynamics of time of arrival operators.

\section*{Data Availability Statement}
There is no data associated with this manuscript.

\section*{Acknowledgment}\label{sec:acknow}
D.A.L. Pablico gratefully acknowledges the support of the Department of Science and Technology - Science Education Institute (DOST-SEI) through the Accelerated Science and Technology Human Resource Development Program (ASTHRDP) graduate schrolarship program.


\bibliography{ref}
\bibliographystyle{unsrt}

\end{document}